\let \emph \textit
 
\documentclass{ieeeaccess}

\usepackage{amsmath,amssymb,amsfonts}

\usepackage{algorithmic}
\usepackage{graphicx}
\usepackage{textcomp}
 \usepackage{multirow}
\usepackage{float} 
\usepackage{subfigure}  

\usepackage{inputenc}
\usepackage{algorithm}
\usepackage{soul} 
\usepackage{color, xcolor} 

\usepackage{float}
\usepackage{booktabs}
\usepackage{array}
\usepackage{cite}

\def\BibTeX{{\mathrm B\kern-.05em{\sc i\kern-.025em b}\kern-.08em
    T\kern-.1667em\lower.7ex\hbox{E}\kern-.125emX}}
\begin{document}
\history{Date of publication xxxx 00, 0000, date of current version xxxx 00, 0000.}
\doi{10.1109/ACCESS.2021.DOI}

\title{ Dynamic Underwater Acoustic Channel Tracking for Correlated  Rapidly Time-varying  Channels }
\author{\uppercase{Qihang Huang}\authorrefmark{1}, 
\uppercase{Wei Li}\authorrefmark{1,2}, \IEEEmembership{Member, IEEE},  \uppercase{Weicheng Zhan}\authorrefmark{1},
\uppercase{Yuhang Wang}\authorrefmark{1} and \uppercase{Rongrong Guo}\authorrefmark{1}}
\address[1]{Harbin Institute of Technology (Shenzhen), China.}
\address[2]{Peng Cheng Laboratory, Shenzhen, China.}
\tfootnote{This work is supported by the National Natural Science Foundation of China (61871151), National Natural Science Foundation of Guangdong Province (2018A030313177), Guangdong Science and Technology Planning Project (2018B030322004), and the project ``The Verification Platform of Multi-tier Coverage Communication Network for oceans (LZC0020)''.}

\markboth
{Qihang Huang\headeretal: Dynamic Underwater Acoustic Channel Tracking for Correlated  Rapidly Time-varying  Channels}
{Qihang Huang \headeretal: Dynamic Underwater Acoustic Channel Tracking for Correlated  Rapidly Time-varying  Channels}

\corresp{Corresponding author: Wei Li (e-mail: li.wei@hit.edu.cn).}

\begin{abstract}
In this work, we focus on the model-mismatch problem for model-based subspace channel tracking in the correlated underwater acoustic channel. A model based on the underwater acoustic channel's correlation can be used as the state-space model in the Kalman filter to improve the underwater acoustic channel tracking compared that without a model. Even though the data support the assumption that the model is slow-varying and uncorrelated to some degree, to improve the tracking performance further, we can not ignore the model-mismatch problem because most channel models encounter this problem in the underwater acoustic channel. Therefore, in this work, we provide a dynamic time-variant state-space model for underwater acoustic channel tracking. This model is tolerant to the slight correlation after decorrelation. Moreover, a forward-backward Kalman filter is combined to further improve the tracking performance. The performance of our proposed algorithm is demonstrated with the same at-sea data as that used for conventional channel tracking. Compared with the conventional  algorithms, the proposed algorithm shows significant improvement, especially in rough sea conditions in which the channels are fast-varying.
\end{abstract}


\titlepgskip=-15pt

\maketitle
\sethlcolor{yellow}
\section{Introduction}
\PARstart{U}{nderwater} acoustic channels are some of the most challenging channels in wireless communications \cite{2002The} because they may suffer from substantial multipath interference, significant Doppler spread, and rapid time variation. 
Acquiring the acoustic channel state information is usually crucial for underwater acoustic communications and signal processing \cite{2020Channel}.
 
With the development of underwater acoustic channel estimation and tracking, there has emerged a trend of researchers increasingly considering the physics of the channel before to search for solutions in radio communications\cite{2019ASong, BZPW10, chen2018underwater, Huang2013Multipath}. One of the physical phenomena  of the underwater acoustic channel was demonstrated in \cite{Huang2013Multipath, Nadakuditi2004A, yang2012properties}, where it was shown that for many underwater acoustic channel impulse responses (CIRs), the multipath taps are correlated. This result may be due to the fluctuations of the water dividing the path into several similar paths. The signals may travel through a common region with similar sound speed variations. Additionally,  the limitation of the channel bandwidth or bandwidth of the receivers and transmitters leads to the tap belonging to one path spreading to multiple adjacent taps \cite{Huang2013Multipath}. 

The correlation characteristic of the underwater acoustic channels may potentially be used to improve the channel tracking since after certain decorrelation, the channel can become uncorrelated and may have a lower rank \cite{1996Tsatsanis}. 
Even though  some artificial intelligence (AI)-based protocols have been developed in recent studies of underwater acoustic communication and networking \cite{Jin2019, Chen2020, Alamgir2020}, they usually incur a high computational cost. 
However,  based on the correlation characteristic,  the performance can be improved and the computational cost can be reduced \cite{1996Tsatsanis, Huang2014Model, Huang2015Improving}. A model-based signal subspace channel tracking approach is  proposed in \cite{Huang2014Model} for the correlated underwater acoustic channel. It is assumed that after decorrelation, the channel components become uncorrelated and slow-varying. A time-invariant autoregressive (AR) model with low order is used as the model for these uncorrelated channel components after decorrelation. This AR model has a simple form and can be used for the time evaluation of the channel components. Some experimental results support the conclusion that this AR model is reasonable for channel components \cite{Huang2014Model, Huang2015Improving}. 
With this time-invariant AR model, a Kalman filter is used to track these channel components, thus realizing correlated underwater acoustic channel tracking. Based on this time-invariant AR model for channel components, adaptive subspace-tracking with reduced-rank model-based amplitude estimation (ASRMAE) in \cite{Huang2015Improving} improves this channel tracking by adding a subspace tracker for the time variation of the subspace vector.  An adaptive factor  is introduced  into the tracker in \cite{wang2019training} to further improve the tracking ability. 

However, model mismatch often happens in model-based underwater acoustic channel tracking since to date, no model has been able to fully describe the underwater acoustic channel. Once the channel components are not correlated and not evaluated in a time-invariant manner, the time-invariant AR model mismatches the channel components. Especially for the rapid time-varying underwater acoustic channel, this prior model mismatch is nonnegligible \cite{Han2020}. 
Thus,  even though improved trackers  can compensate for the mismatch to some degree  \cite{wang2019training, Han2020}, a more tolerant model must be developed.
The model itself needs to be tolerant to the occasionally uncorrelated channel components and time-variant evolutions,  thereby improving the tracking performance in principle.

Based on the above considerations, the main contributions of this work are as follows:
\begin{itemize}

	\item We propose a tolerant dynamic time-variant state-space model for the channel components in channel tracking. In the state-space model, we dynamically update the state-space model during the tracking process. Moreover, the model is tolerant to the model mismatch if the uncorrelated assumption for the channel components does not hold in the nonstationary underwater acoustic channel. Through this approach, we improve the state-space model for the rapid time-varying channel tracking.
	
	\item Moreover, a forward-backward Kalman filter is combined with the dynamic time-variant state-space model for rapid time-varying channel tracking named as dynamic forward-backward ASRMAE (DFB-ASRMAE). Since the rapid time-varying channel requires a high tracking ability, we provide the two-way tracker to improve tracking efficiency based on the proposed state-space model. Our results show that the proposed channel tracking decreases the normalized signal prediction error for rapid time-varying acoustic channels compared to the existing methods with the same experimental data.	
	
\end{itemize}

\section{System Model}
\subsection{ Correlated Underwater Acoustic Channels}

A time-varying underwater acoustic channel  has $M$ multipath arrivals that each have an amplitude ${A_m}(t)$ and arrival (delay) time ${\tau _m}(t)$:
\begin{equation}
c(\tau,t ) = \sum\limits_{m = 1}^M{{A_m}(t)} \delta (\tau  - \tau_m(t)),
\end{equation}
where $c(\tau,t )$  is the channel impulse response at time $t$.
Then, at the receiver, the effective CIR can be obtained: 
\begin{equation}
h(\tau,t) = {p_\mathrm{R}}(\tau ) \otimes c(\tau,t ) \otimes {p_\mathrm{T}}(\tau ),
\end{equation}
where ${p_\mathrm{R}}(\tau)$ and ${p_\mathrm{T}}(\tau )$ are the receive filter and transmit filter, respectively.
After sampling $h(\tau,t)$,
we have $h_k(n) =h(\tau  = k{T_\mathrm{b}} ,t = n{T_\mathrm{g}})$ for the $k$th tap at time $n$ with $n=1,2,...,N$, where ${T_\mathrm{b}}$ is the duration of a symbol, given that the channel is acquired every ${T_\mathrm{g}}$ duration. Then, the CIR can be expressed in terms of a $K$-tap  vector $\mathbf{h}(n) = {[{h_1}(n),\ldots, {h_k}(n), \ldots ,{h_K}(n)]^T}$.
For many underwater acoustic channels, the taps  are found to be experimentally correlated \cite{Huang2013Multipath, Nadakuditi2004A}.
This is reasonable because many physical environments  can make the taps correlated. 
The fluctuations of the water can divide the path into several similar paths. 
When many paths travel through a common region with similar sound speed variations, they may become correlated. Moreover, if  the bandwidths of the receive filter and transmit filter are too narrow, the tap belonging to one path can easily spread to many adjacent taps \cite{Huang2013Multipath}.

The cross-path covariance matrix is given by $\mathbf{R}_h(n) = E[\mathbf{h}(n)\mathbf{h}^H(n)]$ 
where superscript $H$ denotes the Hermitian conjugate. Through eigenvalue decomposition (EVD) of the covariance matrix,      
\begin{equation}\label{EVDR}
\mathbf{R}_h(n)=\mathbf{Q}(n)\mathbf{\Lambda}(n)\mathbf{Q}^H(n),
\end{equation}
where $\mathbf{Q}(n)$ is a matrix of orthonormal eigenvectors, and $\mathbf{\Lambda}(n)$ is  a diagonal matrix of eigenvalues. Since channel taps are correlated, $\mathbf{R}_h(n)$ is  ill-conditioned with assumed $r$-rank, where $r \ll K$. 
If the taps of the channel components in the eigenvector space after decomposition are  uncorrelated  \cite{Huang2014Model} (we analyze this in Section V with experimental data), we have 
\begin{equation}\label{unco}
\begin{aligned}
\mathbf{\Lambda}(n)=E[\mathbf{z}(n)\mathbf{z}^H(n)],
\end{aligned}
\end{equation}
where $\mathbf{z}(n)$ are the channel components.
The channel taps can be expressed as
\begin{equation}\label{h2z}
\begin{aligned}
\mathbf{h}(n) &= {\mathbf{Q}}(n){\mathbf{z}}(n)&\\
&= {{\mathbf{Q}}}_r(n){{\mathbf{z}}}_r(n),&
\end{aligned}
\end{equation}
where $\mathbf{z}_r(n)\in\mathbb{C}^{r\times 1}$  comprises  the $r$ largest channel components in 
${{\mathbf{z}}}(n)$, and the eigenvectors associated with the $r$ largest eigenvalues  compose ${{\mathbf{Q}}}_r(n)\in\mathbb{C}^{K\times r}$ where 
${\mathbf{Q}_r}^H(n){\mathbf{Q}_r}(n)={\mathbf{I}}_{r \times r}$.

It is obvious that if we can obtain ${\mathbf{Q}}_r(n)$ and ${{\mathbf{z}}}_r(n)$ at each state, then the underwater acoustic channel $\mathbf{h}(n)$ can be tracked.
Compared to $\mathbf{Q}_r(n)$, the channel principal components ${\mathbf{z}}_r(n)$ vary rapidly over time.  
To track  the channel $\mathbf{h}(n)$, it is  more important to track ${\mathbf{z}}_r(n)$. 
Therefore, in this work, we focus on the tracking of channel components ${\mathbf{z}}_r(n)$,
with the tracked the channel subspace ${\mathbf{Q}}_r(n)$ according to \cite{Huang2015Improving}.

\subsection{AR Model to the Channel  Principle Components }
To track the channel, we can use an AR model to describe the time evolution of the signal components $\mathbf{z}_{r}(n)$. Based on this model, the signal components $\mathbf{z}_{r}(n)$ are considered to follow a $p$-order discrete-time Markov process \cite{tsatsanis1996estimation}.
\begin{equation}\label{ARm}
\mathbf{z}_{r}(n)=\sum_{l=1}^{p} \boldsymbol{\Phi}(n, l) \mathbf{z}_{r}(n-l)+\boldsymbol{\eta}(n),
\end{equation}
where $\boldsymbol{\Phi}(n, l)$ is the $l$th-order  state transition matrix  at time $n$, and $\boldsymbol{\eta}(n)$ represents the process noise vector at time $n$ and $\boldsymbol{\eta}(n)=\left[\eta_{1}(n), \eta_{2}(n), \ldots, \eta_{r}(n)\right]^{T}$. Multiplying \eqref{ARm} by $\mathbf{z}_{r}^{H}(n-m)$ where $0\leq m \leq  p$  from the right side and taking expectation,  we obtain 
\begin{equation}\label{AR}
\begin{split}
E\left[\mathbf{z}_{r}(n)\mathbf{z}_{r}^{H}(n-m)\right]=E\left[\boldsymbol{\eta}(n)\mathbf{z}_{r}^{H}(n-m)\right]\\
+\sum_{l=1}^{p} \boldsymbol{\Phi}(n, l) \times E\left[\mathbf{z}_{r}(n-l) \mathbf{z}_{r}^{H}(n-m)\right].
\end{split}
\end{equation}
If the taps of the signal components are uncorrelated,  $E\left[z_{i}(n) z_{j}^{*}(l)\right]=R_{i}(n-l) \delta_{i j}$, where $z_{i}(n)$ represents the $i$th tap of $\mathbf{z}_{r}(n)$, $R_{i}(n-l)=E\left[z_{i}(n) z_{i}^{*}(l)\right]$ and  is the autocorrelation  of tap $i$ component. Given that $\eta_{i}(n)$ and $z_{j}(n-m)$ are independent random variables, 
\begin{equation}\label{cnz}
\begin{aligned}
E\left[\eta_{i}(n) z_{i}^{*}(n-m)\right]=& \sum^p_{l=1}{\Phi}_{i}(n, l)E\left[\eta_{i}(n) z_{i}^{*}(n-m-l)\right]\\
&+E\left[\eta_{i}(n) \eta_{i}^{*}(n-m)\right]\\
=&R_{\eta, i} \delta(m),
\end{aligned}
\end{equation}
 where $R_{\eta, i}\!\!\!=\!\!\!E\left[\eta_{i}(n) \eta_{i}^{*}(n)\right]$ and is the process noise variance corresponding to the $i$th tap.
Thus,  \eqref{AR} for the $i$th tap can be expressed as \cite{Huang2014Model, Huang2015Improving}
\begin{equation}\label{R_im}
R_{i}(m)\!=\!\sum_{l=1}^{p} {\Phi}_{i}(n, l)R_{i}(m-l)+R_{\eta, i} \delta(m) \quad i \!= \!1,2, \ldots, r.
\end{equation}
The Yule-Walker equation can help us find the AR coefficients ${\Phi}_{i}(n, l)$.

\section{Conventional ASRMAE Subspace Channel Tracking }

The following conventional ASRMAE subspace channel tracking is taken from \cite{Huang2014Model, Huang2015Improving, wang2019training}.
In conventional subspace channel tracking, the state transition coefficient ${\Phi}_{i}(n, l)={\phi}_{i}(l)$ and is assumed to be time invariant.
The channel principal components are modeled as an AR process \eqref{ARm}, and a Kalman filter recursively tracks the channel components based on this AR model.
The state-space and the observation models in the Kalman filter are as follows:
\begin{equation}\label{Deq}
\mathbf{Z}(n)=\mathbf{\Phi}_{z} \mathbf{Z}(n-1)+\boldsymbol{\eta}_{*}(n),
\end{equation}
\begin{equation}\label{Meq}
r(n)=\mathbf{D}(n) \mathbf{Z}(n)+v(n),
\end{equation}
where  $\mathbf{Z}(n)=\left[\mathbf{z}_{r}^{T}(n),  \ldots, \mathbf{z}_{r}^{T}(n-p+1)\right]^{T}$, $\boldsymbol{\eta}_{*}(n)=\left[\boldsymbol{\eta}^{T}(n), \boldsymbol{0}_{1 \times r(p-1)}\right]^{T}$ and is the process noise vector assumed to be uncorrelated with each tap,  and the state transition matrix $\mathbf{\Phi}_{z}$ is denoted by
\begin{equation}\label{phiz}
\mathbf{\Phi}_{z}=\left[\begin{matrix}
\mathbf{\Phi}(1) & \mathbf{\Phi}(2) & \cdots & \mathbf{\Phi}(p) \\
\mathbf{I}_{r} & \mathbf{0}_{r} & \cdots & \mathbf{0}_{r} \\
\vdots & \ddots & \ddots & \vdots \\
\mathbf{0}_{r} & \cdots & \mathbf{I}_{r} & \mathbf{0}_{r}
\end{matrix}\right],
\end{equation}
where $\mathbf{\Phi}(1)=\operatorname{diag}\left(\left[\phi_{1}(1), \phi_{2}(1), \ldots, \phi_{r}(1)\right]\right)$ and is the first-order state transition matrix with time-invariant assumption. 
The state-space model \eqref{Deq} in the Kalman filter is obtained by a rearranged \eqref{ARm} with time-invariant state transition coefficients.
In the observation model \eqref{Meq}, the received signal $r(n)$ can be expressed as $r(n)=\mathbf{d}(n)^{T} \mathbf{h}(n)+v(n)$, where $\mathbf{d}(n)=[d(n), d(n-1), \ldots d(n-K)]^{T}$ and is the transmitted data vector, and $v(n)$ is the observation noise, which is additive Gaussian white noise. 
 $\mathbf{d}_{z}(n)^{T}=\mathbf{d}(n)^{T} \mathbf{Q}_{r}(n)$, and is the projection of the transmitted data vector on the signal subspace. 
Additionally, $\mathbf{D}(n)=\left[\mathbf{d}_{z}(n)^{T}, \mathbf{0}_{1 \times r(p-1)}\right]$.

To achieve the Kalman filter with  \eqref{Deq} and \eqref{Meq} to track the channel components $\mathbf{z}_r(n)$, we 
  need to estimate the state transition matrix $\mathbf{\Phi}_{z}$ and process noise variance $\mathbf{R_{\eta}}$ \cite{Huang2015Improving}.  
\subsection{Estimation of State Transition Matrix $\mathbf{\Phi}_{z}$}\label{CSTM}

Since the state transition coefficients are assumed to be time invariant,
we convert $\left\{R_{i}(m)\right\}_{m=1}^{p}$ into a vector form as $\mathbf{R}_{i}=\left[R_{i}(1), R_{i}(2), \ldots, R_{i}(p)\right]^{T}$, then \eqref{R_im} can be transformed into
\begin{equation}\label{YWe}
\mathbf{R}_{i}  \!\!\!=  \!\!\!\underbrace{\left[\begin{matrix}
	R_{i}(0) \!\!\!&R_{i}(-1) \!\!\!&\cdots \!\!\!&R_{i}(-p+1) \\
	R_{i}(1) \!\!\!&R_{i}(0) \!\!\!&\cdots \!\!\!&R_{i}(-p+2) \\
	\vdots \!\!\!&\ddots \!\!\!&\ddots \!\!\!&\vdots \\
	R_{i}(p-1) \!\!\!&R_{i}(p-2) \!\!\!&\cdots \!\!\!&R_{i}(0)
	\end{matrix}\right]}_{\mathbf{r}_{i}}\!\!\!\times\!\!\!\underbrace{\left[\begin{matrix}
	\phi_{i}(1) \\
	\phi_{i}(2) \\
	\vdots \\
	\phi_{i}(p)
	\end{matrix}\right]}_{\mathbf{\Phi}_{i}}.    
\end{equation}
\eqref{YWe} is known as the Yule-Walker equation, and then, the state transition coefficients for the $i$th tap  can be obtained as 
\begin{equation}\label{phii}
\mathbf{\Phi}_{i}=\mathbf{r}_{i}^{-1} \mathbf{R}_{i}. 
\end{equation}
It is clear that the state transition matrix $\mathbf{\Phi}_{z}$ in \eqref{Deq} can be obtained  by reshaping $\mathbf{\Phi}_{i}$ as \eqref{phiz}. 

According to \eqref{YWe} and \eqref{phii}, to  estimate $\mathbf{\Phi}_{i}$, it is necessary to first estimate ${R}_{i}(m)$. However, due to the lack of prior information about the channel principle components,
we resort to using the least mean square (LMS) method for a coarse estimation of the CIR in \eqref{h2z}, and then, we can obtain the channel components  $\mathbf{z}_r(n)$.

We denote $\hat{\mathbf{h}}^{\mathrm{LMS}}(n)$ as the estimated CIR at time $n$ by LMS.  The estimation error in LMS can be expressed as
$e(n)=r(n)- (\hat{\mathbf{h}}^{\mathrm{LMS}})^{H}(n) \boldsymbol{d}(n)$.
Then, iteratively, we can roughly estimate the channel by LMS as follows:
\begin{equation}
\hat{\mathbf{h}}^{\mathrm{LMS}}(n+1)=\hat{\mathbf{h}}^{\mathrm{LMS}}(n)+2 \mu e(n) \boldsymbol{d}(n),
\end{equation}
where $\mu$ is the step factor.
The estimated time-invariant cross-correlation matrix of channel is given by
\begin{equation}
\hat{\mathbf{R}}_{h}=\frac{1}{N_{p}} \sum_{n=1}^{N_{p}} \hat{\mathbf{h}}^{\mathrm{LMS}}(n) (\hat{\mathbf{h}}^{\mathrm{LMS}})^{H}(n),
\end{equation}
where $N_{p}$ is the length of the training sequence. Through EVD as described in (\ref{EVDR}), we can obtain the eigenvectors $\hat{\mathbf{Q}}_{r}(n)$ corresponding to  the $r$ largest eigenvalues.
A rough estimation of the channel components can be obtained by projecting the $\hat{\mathbf{h}}^{\mathrm{LMS}}(n)$ onto $\hat{\mathbf{Q}}_{r}(n)$ according to (\ref{h2z}),
$\hat{\mathbf{z}}_{r}^{\mathrm{LMS}}(n)=\hat{\mathbf{Q}}_{r}^{H}(n) \hat{\mathbf{h}}^{\mathrm{LMS}}(n)$.    
With  $\hat{\mathbf{z}}_{r}^{{\mathrm{LMS}}}(n)$, we have
\begin{equation}\label{estR_im}
\hat{R}_{i}(m)=\frac{1}{N_{p}} \sum_{n=1}^{N_{p}} \hat{z}_{i}^{\mathrm{LMS}}(n)  (\hat{z}_{i}^{\mathrm{LMS}})^* (n-m).
\end{equation}

After we obtain $\hat{R}_{i}(m)$, we can find $\mathbf{R}_{i}$ and ${\mathbf{r}_{i}}$ as in \eqref{YWe}. Then, with  \eqref{phii}, $\mathbf{\Phi}_{i}$ can be estimated. Therefore, we can arrive at the state transition matrix $\mathbf{\Phi}_{z}$.

\subsection{Estimation of Process Noise Variance $\mathbf{R_{\eta}}$}
The conventional estimation of $\mathbf{R_{\eta}}$  is based on the assumption that the channel components of different taps $\mathbf{z}(n)$  are uncorrelated.

Letting $m=0$ in \eqref{R_im}, with the time-invariant state transition coefficient ${\phi}_{i}(l)$, we have 
\begin{equation}\label{R0}
R_{i}(0)\!=\!\sum_{l=1}^{p} {\phi}_{i}(l)R_{i}(0-l)+R_{\eta, i} ,\quad \quad i \!= \!1,2, \ldots, r.
\end{equation}
Since ${R}_{i}(m)$ can be estimated from \eqref{estR_im} and ${\phi}_{i}(l)$ can be obtained from \eqref{phii},
 $R_{\eta, i}=R_{i}(0)-\sum_{l=1}^{p} {\phi}_{i}(l)R_{i}(0-l)$. 

By assuming  that the channel  components of different taps $\mathbf{z}(n)$ are uncorrelated as in \eqref{unco}, 
the left and right  sides of  \eqref{AR} should be diagonal matrices, and the process noise is uncorrelated with different taps.
Then,  according to \eqref{AR}, \eqref{cnz}  and \eqref{R0}, the  noise variance matrix $\mathbf{R_{\eta}}$ is a diagonal matrix  and can be expressed as
\begin{equation}\label{Rn}
\mathbf{R_{\eta}}=\operatorname{diag}\left(\left[R_{\eta, 1}, R_{\eta, 2}, \ldots, R_{\eta, r}\right]\right).
\end{equation}

\section{Dynamic Forward-backward Subspace Channel Tracking for a Rapidly Time-varying Underwater Acoustic Channel}

In this section, we propose a dynamic forward-backward subspace channel tracking method. We first provide a time-variant state-space model. 
In the time-variant state-space model, we  update the state transition matrix dynamically and provide a new estimation of process noise statistics. The model mismatch  due to the fast time-varying channel can be effectively mitigated. Then, to further adjust the time-varying property, a forward-backward Kalman filter is combined with the dynamic state-space model as the proposed DFB-ASRMAE.

\subsection{Dynamic Model Estimation for a Rapidly Time-varying Channel}

The proposed dynamic state-space model is as follows:
\begin{equation}\label{Daeq}
\mathbf{Z}(n)=\mathbf{\Phi}_{z}(n) \mathbf{Z}(n-1)+\boldsymbol{\eta}_{*}(n).
\end{equation}
Different from \eqref{Deq},  we assume that $\mathbf{\Phi}_{z}(n)$ changes with time $n$  in \eqref{Daeq}, and $\boldsymbol{\eta}_{*}(n)$ is correlated with taps due to the model-mismatch in the rapidly time-varying channel. Thus, we can still use a low-order  AR model to adapt to the rapidly time-varying channel.

\subsubsection{Dynamically Updating the State Transition Matrix for a Rapidly Time-varying Channel}

Here, we introduce a dynamically updated state transition matrix ${\boldsymbol{\Phi}}_{z}(n)$ in \eqref{Daeq}  based on  the state-space model \eqref{Daeq} and \eqref{Meq}.

We first obtain the initial ${\boldsymbol{\Phi}}_{z}(0)={\boldsymbol{\Phi}}_{z}$ with the conventional method as in Section III. A. Then, we use the following algorithm to update the state transition matrix every $T_{\mathrm g}$ duration based on the Kalman filter.
The Kalman filter includes two main steps: update and prediction  \cite{Kalman1960Approach}.

 \begin{itemize}
 	\item \textbf{Update}
 \end{itemize}
 \begin{equation}\label{U1}
 g(n)=\mathbf{D}(n) \hat{\mathbf{K}}(n,n-1) \mathbf{D}^{H}(n)+\sigma_{v}^{2},
 \end{equation}
 \begin{equation}\label{U2}
 \mathbf{G}(n)= \hat{\mathbf{K}}(n,n-1) \mathbf{D}^{H}(n)g(n)^{-1},
 \end{equation}
 \begin{equation}\label{U3}
 \xi(n)=r(n)-\mathbf{D}(n) \hat{\mathbf{Z}}(n,n-1),
 \end{equation}
 \begin{equation}\label{zn}
 \hat{\mathbf{Z}}(n,n)= \hat{\mathbf{Z}}(n,n-1)+\mathbf{G}(n)\xi(n),
 \end{equation}
 \begin{equation}\label{U4}
 \hat{\mathbf{K}}(n,n)=\hat{\mathbf{K}}(n,n-1)- \mathbf{G}(n) \mathbf{D}(n) \hat{\mathbf{K}}(n,n-1),
 \end{equation}
 \begin{itemize}
 	
 	\item \textbf{Prediction }
 	
 \end{itemize}
 \begin{equation}\label{P1}
 \hat{\mathbf{Z}}(n+1,n)={\boldsymbol{\Phi}}_{z}(n) \hat{\mathbf{Z}}(n, n),
 \end{equation}
 \begin{equation}\label{P2}
 \hat{\mathbf{K}}(n+1,n)={\boldsymbol{\Phi}}_{z}(n) \hat{\mathbf{K}}(n, n) {\boldsymbol{\Phi}}_{z}^{H}(n)+\mathbf{R_{\eta}},
 \end{equation}
 where $\hat{\mathbf{Z}}(n,n)=\left[\hat{\mathbf{z}}_{r}^{T}(n),  \ldots, \hat{\mathbf{z}}_{r}^{T}(n-p+1)\right]^{T}$ and is the estimated state vector with the estimated channel components, $\sigma_{v}^{2}$ represents the variance of the observation noise, and ${\boldsymbol{\Phi}}_{z}(n)$ is the state transition matrix at time $n$, which changes with time.  In the update step, given the state vector  $\hat{\mathbf{Z}}(n, n-1)$ and the Kalman error covariance matrix $\hat{\mathbf{K}}(n, n-1)$ predicted at the previous moment, we can obtain the Kalman gain $\mathbf{G}(n)$ and the signal prediction error  $\xi(n)$. In the prediction step, the predicted state vector $\hat{\mathbf{Z}}(n+1, n)$ and predicted covariance matrix $\hat{\mathbf{K}}(n+1,n)$ are obtained.

To predict the state transition matrix ${\boldsymbol{\Phi}}_{z}(n+1)$,  we extract $\hat{{z}_{i}}(n+1)$ from the predicted  state vector $\hat{\mathbf{Z}}(n+1,n)$.
The  predicted channel component autocorrelation for the $i$th tap is as follows:
 \begin{align}\label{Dst}
 \hat{R}_{i}(m,n+1)=&\frac{1}{n+1} \sum_{l=1}^{n+1} \hat{z}_{i}(l) \hat{z}_{i}^{*}(l-m)\nonumber\\
 =&\frac{n}{n+1} \hat{R}_{i}(m,n)\nonumber\\
 &+ \frac{1}{n+1} \hat{z}_{i}(n+1) \hat{z}_{i}^{*}(n+1-m),
 \end{align}
 where $\hat{R}_{i}(m,n)$ is time variant $\hat{R}_{i}(m)$, and  $\hat{R}_{i}(m,n)$ is estimated with the data obtained before time $n$, rather than $N_p$ as used in \eqref{estR_im} for $\hat{R}_{i}(m)$.
With  $\hat{R}_{i}(m,n+1)$, we can obtain  the predicted state transition matrix ${\boldsymbol{\Phi}}_{z}(n+1)$ according to \eqref{phii}.

For many underwater acoustic channels, especially rapidly time-varying channels, the model time variation is not negligible. The conventional time-invariant state transition matrix can lead to  accumulative mismatch for the state-space model. By contrast, the proposed dynamically updated state transition matrix can lower such mismatch.

\subsubsection{Correlated  Process Noise Covariance Estimation}

For the conventional state-space model, it is assumed that after  EVD,  the subspaces corresponding to the channel components are  orthogonal to each other, and the channel components are uncorrelated;
therefore, the noise in state-space should be uncorrelated.
However, for the fast-varying channel, after decorrelation,  the channel components may still be slightly correlated since the channel changes quickly.
 Then, the process noise becomes correlated. The diagonal assumption about the $\mathbf{R_{\eta}}$ in (\ref{Rn}) no longer holds.

With the conventional method in Section III. A, we can obtain $\boldsymbol{\Phi}_z(0)$ and
the rough channel  components $\hat{\mathbf{z}}_{r}^{\mathrm{LMS}}(n)$. 
Then, with $\hat{\boldsymbol{\Phi}}(l)$ extracted from $\boldsymbol{\Phi}_z(0)$  and  $\hat{\mathbf{z}}_{r}^{\mathrm{LMS}}(n)$, we can estimate the process noise vector based on \eqref{ARm} as follows:
\begin{equation}\label{pn}
\hat{\boldsymbol{\eta}}(n)=\hat{\mathbf{z}}_{r}^{\mathrm{LMS}}(n)-\sum_{l=1}^{p} \hat{\boldsymbol{\Phi}}(l) \hat{\mathbf{z}}_{r}^{\mathrm{LMS}}(n-l).
\end{equation} 
Then, for the correlated process noise, the estimated  covariance of the process noise $\hat{\mathbf{R}}_{\eta}$ is given by
\begin{equation}\label{cpn}
\hat{\mathbf{R}}_{\eta}=\frac{1}{N_{p}} \sum_{n=1}^{N_{p}} \hat{\boldsymbol{\eta}}(n) \hat{\boldsymbol{\eta}}^{H}(n),
\end{equation}
where $N_{p}$ is the time-length of the training data.
Compared with the conventional method, $\hat{\mathbf{R}}_{\eta}$ is no longer a diagonal matrix. It can help the state-space model to adapt to the fast-varying underwater acoustic channel tracking. $\hat{\mathbf{R}}_{\eta}$ does not need to be updated at every time state.

\subsection{ Forward-Backward Kalman Filter for Channel Tracking}

 In this section, 
to further improve the channel tracking, we combine the dynamically updated state-space model proposed above with forward-backward Kalman filtering \cite{Murwan2018kalman, 2020New}  as DFB-ASRMAE. 
The forward-backward Kalman filter can better adapt to the time-variant state-space model.
The forward-backward Kalman filtering includes three steps: forward filtering, backward filtering, and optimal joint estimation.

In the backward filtering, different from the conventional Kalman filter tracking along time $n=1$ to $N$ as the forward tracking, the backward Kalman filter recursively tracks from time $n=N$ back to $1$. 
Since $\mathbf{\Phi}(n)$ is a nonsingular diagonal matrix, multiplying the conventional state model (\ref{Daeq}) by $\mathbf{\Phi}^{-1}(n)$ on the left, we have:
\begin{equation}
\mathbf{\Phi}^{-1}(n) \mathbf{Z}(n)= \mathbf{Z}(n-1)+\mathbf{\Phi}^{-1}(n) \boldsymbol{\eta}(n).
\end{equation}
Then, we have the backward Kalman filter with the backward  state-space model and the observation model as follows
\begin{equation}\label{db}
\mathbf{Z}(n-1)=\mathbf{\Phi}_{\mathrm b}(n) \mathbf{Z}(n)+\boldsymbol{\eta}_{\mathrm b}(n),
\end{equation}
\begin{equation}\label{zn}
 {r}(n)=\mathbf{D}(n) \mathbf{Z}(n)+{v}(n),
\end{equation}
where subscript $\mathrm b$ represents the parameters for the backward Kalman filter. Clearly, the state transition matrix and the process noise can be obtained as $\mathbf{\Phi}_{\mathrm b}(n)=\mathbf{\Phi}^{-1}(n)$ and $\boldsymbol{\eta}_{\mathrm b}(n)=\mathbf{\Phi}^{-1}(n) \boldsymbol{\eta}(n)$, respectively. Then, backward Kalman filtering can similarly apply the sequential procedure in Kalman filtering, from time $n=N$ back to $1$.

To explain the forward-backward Kalman filter for the channel tracking, here, we simplify the forward system \eqref{Daeq}\eqref{Meq}
and the backward system \eqref{db}\eqref{zn} as follows:
\begin{equation}\label{zf}
{r}_{\mathrm f}(n)=\mathbf{D}_{\mathrm f}(n) \mathbf{Z}(n)+{v}_{\mathrm f}(n),
\end{equation}
\begin{equation}\label{zb}
 {r}_{\mathrm b}(n)=\mathbf{D}_{\mathrm b}(n) \mathbf{Z}(n)+{v}_{\mathrm b}(n),
\end{equation}
where $\mathbf{Z}(n)$ is to be estimated, ${r}_{\mathrm f}(n)$ and ${r}_{\mathrm b}(n)$ are the outputs of the forward and backward systems, $\mathbf{D}_{\mathrm f}(n)$ and $\mathbf{D}_{\mathrm b}(n)$ represent the system matrix, and ${v}_{\mathrm f}(n)$ and ${v}_{\mathrm b}(n)$ represent the observation noise vectors. Since Kalman filtering is a generalization of sequential linear minimum mean square  estimation (LMMSE), 
we study the LMMSE of $\hat{ \mathbf{Z}}_{\mathrm f}(n)$ for the forward system:
\begin{equation}\label{Zf}
\begin{aligned}
\hat{ \mathbf{Z}}_{\mathrm f}(n)=&\left(\mathbf{R}_{\mathbf{z}}^{-1}+{\mathbf{D}_{\mathrm f}}^{H}(n) \mathbf{D}_{\mathrm f}(n) /\sigma_{v{\mathrm f}}^{2}\right)^{-1} \mathbf{D}_{\mathrm f}^{H}(n)  {r}_{\mathrm f}(n) /\sigma_{v{\mathrm f}}^{2}\\
=&\mathbf{M}_{\mathrm f}(n) \mathbf{D}_{\mathrm f}^{H}(n)  {r}_{\mathrm f}(n) /\sigma_{v{\mathrm f}}^{2},
\end{aligned}
\end{equation}
where $\mathbf{R}_{\mathbf{z}}$ is the covariance  of $\mathbf{Z}(n)$, and $\sigma_{v{\mathrm f}}^{2}$ is the variance of $v_{\mathrm f}(n)$. $\mathbf{M}_{\mathrm  f}(n)$ represents the estimation error matrix of $\hat{ \mathbf{Z}}_{\mathrm f}(n)$. Similar to \eqref{Zf}, the LMMSE of $\hat{ \mathbf{Z}}_{\mathrm b}(n)$ obtained from the backward system can be expressed as
\begin{equation}\label{Zb}
\begin{aligned}
\hat{ \mathbf{Z}}_{\mathrm b}(n)
=\mathbf{M}_{\mathrm b}(n) \mathbf{D}_{\mathrm b}^{H}(n)  {r}_{\mathrm b}(n) /\sigma_{v{\mathrm b}}^{2}.
\end{aligned}
\end{equation}
Combining the linear system \eqref{zf} with \eqref{zb}, we have
\begin{equation}\label{zc}
 \begin{array}{c}\underbrace{\left[\begin{array}{l}
{r}_{\mathrm f}(n) \\
{r}_{\mathrm b}(n)
\end{array}\right]}\\
\tilde{\mathbf{ r}}(n)\end{array}
 \begin{array}{c}=\underbrace{\left[\begin{array}{l}
\mathbf{D}_{\mathrm f}(n) \\
\mathbf{D}_{\mathrm b}(n)
\end{array}\right]}\mathbf{Z}(n)\\ 
\tilde{\mathbf{D}}(n)\end{array}\begin{array}{c}+\underbrace{\left[\begin{array}{l}
{v}_{\mathrm f}(n)\\
{v}_{\mathrm b}(n)
\end{array}\right]},\\ \tilde{\mathbf{ v}}(n)\end{array}
\end{equation}
where  $\mathbf{C}_{\tilde{\mathbf{ v}}}=\operatorname{diag}\left(\left[\sigma_{v{\mathrm f}}^{2}, \sigma_{v{\mathrm b}}^{2}\right]\right)$ and is the covariance of $\tilde{\mathbf{ v}}(n)$.
In estimation theory, the estimation can be improved with more informational data \cite{Kay1998}.
While the forward and backward systems are not fully independent, the combined system provides more information than either the forward system or backward system only. Therefore, the estimation from the combined approach \eqref{zc}  should be better than that from simple forward or backward filtering. 
The LMMSE of $\mathbf{Z}(n)$ for \eqref{zc} is
\begin{equation}
\begin{aligned}
\widetilde{\mathbf{Z}}(n) =&(\boldsymbol{R}_{\mathbf{z}}^{-1}+ \tilde{\mathbf{D}}^H(n) \mathbf{C}^{-1}_{\tilde{\mathbf{ v}}}\tilde{\mathbf{D}}(n) 
\tilde{\mathbf{D}}^H(n)\mathbf{C}^{-1}_{\tilde{\mathbf{ v}}}  \tilde{\mathbf{r}}(n) \\
=&\left(\mathbf{R}_{\mathbf{z}}^{-1}+
{\mathbf{D}_{\mathrm f}}^{H}(n) \mathbf{D}_{\mathrm f}(n)/\sigma_{v{\mathrm f}}^{2}+
{\mathbf{D}_{\mathrm b}}^{H}(n) \mathbf{D}_{\mathrm b}(n) /\sigma_{v{\mathrm b}}^{2}
\right)^{-1}\\
&\times\left( \mathbf{D}_{\mathrm f}^{H}(n)  {r}_{\mathrm f}(n) /\sigma_{v{\mathrm f}}^{2}
+\mathbf{D}_{\mathrm b}^{H}(n)  {r}_{\mathrm b}(n) /\sigma_{v{\mathrm b}}^{2}
\right)\\
=&\left(\mathbf{M}_{\mathrm f}^{-1}(n) +\mathbf{M}_{\mathrm b}^{-1}(n) -\mathbf{R}_{\mathbf{z}}^{-1}\right)^{-1}\\
&\times
\left( \mathbf{D}_{\mathrm f}^{H}(n)  {r}_{\mathrm f}(n) /\sigma_{v{\mathrm f}}^{2}
+\mathbf{D}_{\mathrm b}^{H}(n)  {r}_{\mathrm b}(n) /\sigma_{v{\mathrm b}}^{2} 
\right)\\
=&\tilde{\mathbf{M}}(n)
\left( \mathbf{D}_{\mathrm f}^{H}(n)  {r}_{\mathrm f}(n) /\sigma_{v{\mathrm f}}^{2}
+\mathbf{D}_{\mathrm b}^{H}(n)  {r}_{\mathrm b}(n)  /\sigma_{v{\mathrm b}}^{2}
\right). 
\end{aligned}
\end{equation}

\begin{table*}[!htbp]
	 \caption{Channel Tracking Algorithm DFB-ASRMAE.}
	\centering
	\begin{tabular}{lr}
		\hline
		Initialization  & No. of operations\\
		\quad Obtain the  rough channel estimation ${\hat{\mathbf{h}}}^{\mathrm{LMS}}(n)$ by LMS;&\\
		 \quad Let ${{\mathbf{z}}}_r^{\mathrm{LMS}}(n)=\mathbf{Q}_{r}^{H}(n){\hat{\mathbf{h}}}^{\mathrm{LMS}}(n)$;& \\
		\quad Estimate ${\mathbf{\Phi}_z(0)}$ according to Section III. A and set the initial value of ${\sigma_{v{\mathrm f}} ^2}$, ${\sigma_{v{\mathrm b}} ^2}$; // Initialize the state transition matrix  &\\
		\quad Estimate the process noise covariance $\hat{\mathbf{R}}_{\eta }$ according to \eqref{pn}\eqref{cpn}; //Estimate the process noise statistics & \\
		For $n=1,2,...,N$  //Forward tracking&\\	
		\quad Track signal subspace $ \mathbf{Q}_r(n)$ by PASTd \cite{1995Projection};& $O(4Kr)$\\	
		\quad \eqref{U1}\eqref{U2}\eqref{U3}\eqref{zn}\eqref{U4} to obtain $\hat{ \mathbf{Z}}(n, n)$ and $\mathbf{K}(n, n)$;// Update in forward Kalman filter&\\
		\quad \eqref{P1}\eqref{P2} to obtain $\hat{ \mathbf{Z}}(n+1, n)$ and $\mathbf{K}(n+1, n);$//Prediction in the forward Kalman filter &$O(3(rp)^3)$\\
    	\quad if $n<N_p$ // Predict $\mathbf{\Phi}(n)$ &\\
		\quad \quad $\mathbf{\Phi}_z(n+1)={\mathbf\Phi}_z(0)$;&\\
		\quad else&\\ 
		\quad \quad Extract ${\hat{z}_i}(n+1)$ from $\hat{\mathbf{Z}}(n+1, n)$ and obtain $\hat{R}_{i}(m, n+1)$ by  \eqref{Dst};&$O(p)$\\
		\quad \quad Substitute $\hat{R}_{i}(m, n+1)$ into  \eqref{phii} to estimate $\mathbf{\Phi}_i(n+1)$ and then obtain  $\mathbf{\Phi}(n+1)$;&$O(p^2+p^3)$\\
		\quad end &\\
		End& \\
		For $n=N,N-1,...,1$ // Backward tracking&\\ 
		\quad Set ${\mathbf\Phi}_{\mathrm b}(0)={\mathbf\Phi}_z^{-1}(0), {\mathbf\Phi}_{\mathrm b}(n)= {\mathbf\Phi}_z^{-1}(n)$ and $\hat{\mathbf{R}}_{\eta \mathrm{b}}={\mathbf\Phi}_{\mathrm b}(N)\hat{\mathbf{R}}_{\eta} {\mathbf\Phi}_{\mathrm b}(N)^H$; & $O((rp)^3)$\\
		\quad  Update the backward Kalman filter to obtain $\hat{ \mathbf{Z}}_{\mathrm b}(n, n)$ and $\mathbf{K}_{\mathrm b}(n, n)$;&\\
                 \quad Prediction in the backward Kalman Filter  to obtain $\hat{ \mathbf{Z}}_{\mathrm b}(n-1, n)$ and $\mathbf{K}_{\mathrm b}(n-1, n)$;&$O(3(rp)^3)$\\			
		End &\\	
		Tracking output &\\	
		\quad \eqref{Mbf} to obtain $\tilde{\mathbf{M}}(n , n)$ with $\mathbf{K}(n , n)$ and $\mathbf{K}_{\mathrm b}(n , n)$;&$O(3(rp)^3)$\\
				\quad \eqref{bftra} to obtain combining optimal $\tilde{\mathbf{Z}}(n)$ with $\mathbf{K}(n , n)$, $\mathbf{K}_{\mathrm b}(n , n)$ and $\hat{\mathbf{Z}}(n , n)$, $\hat{\mathbf{Z}}_{\mathrm b}(n ,n)$;&$O(3(rp)^2)$\\
		\quad $\mathbf{\hat h}_{\mathrm{DFB-ASRMAE}}(n) = {\mathbf{Q}_r}(n)\tilde{\mathbf{Z}}(n)$;\quad //Tracked channel with  DFB-ASRMAE&$O(Kr)$  \\
		\quad $\xi_{\mathrm{DFB-ASRMAE}}(n)=r(n)-\mathbf{D}(n) \tilde{\mathbf{Z}}(n)$.  //Signal prediction error&\\
		End&\\
		\hline
		Total complexity:$O(\max(6Kr,10(rp)^3))$&
	\end{tabular}

	\label{flow}
\end{table*}

\begin{figure*}[h]
\xdef\xfigwd{\textwidth}
\centering
	\subfigure[]{
		\centering
		\includegraphics[scale=0.5]{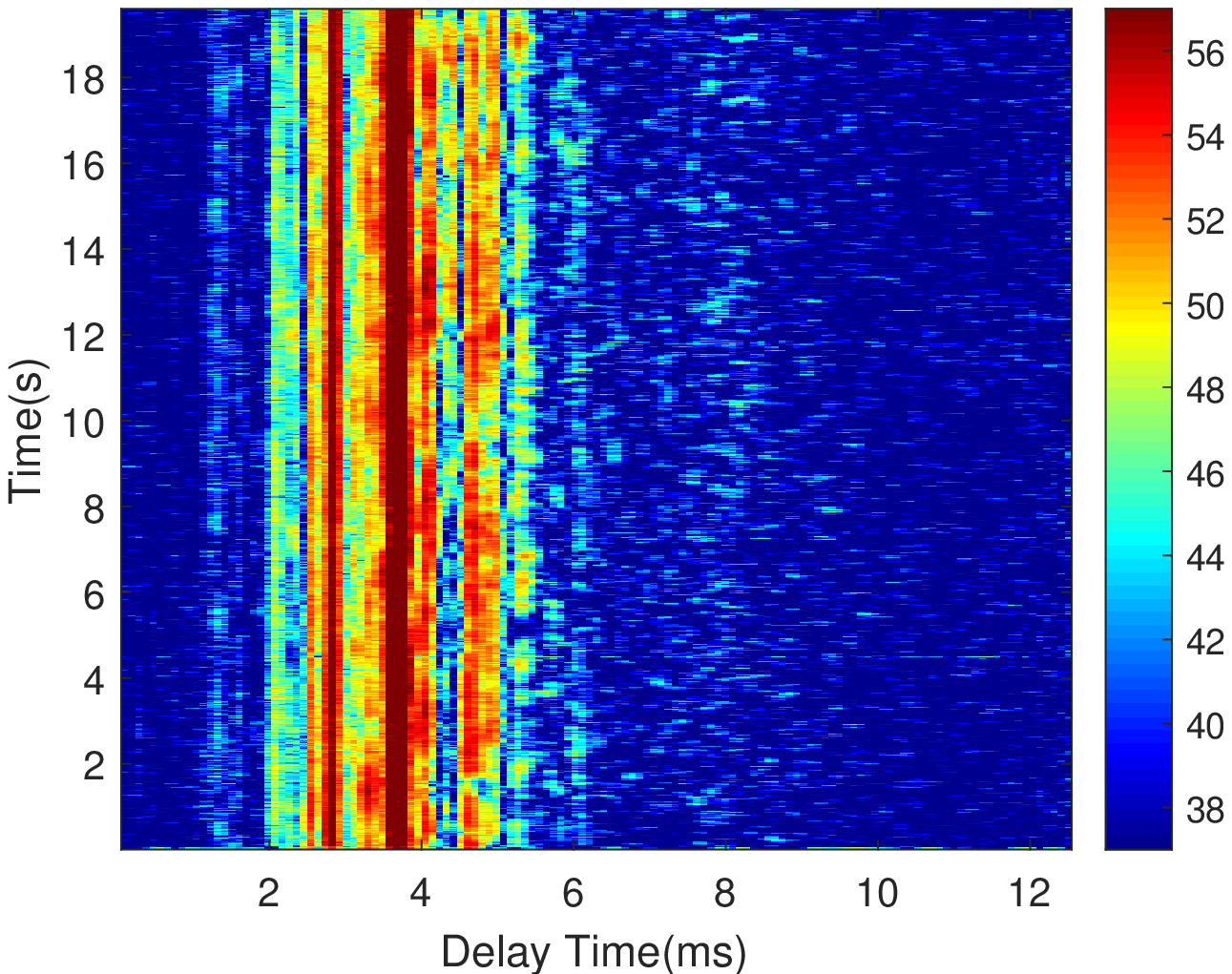}
	}
	\subfigure[]{
		\centering
		\includegraphics[scale=0.5]{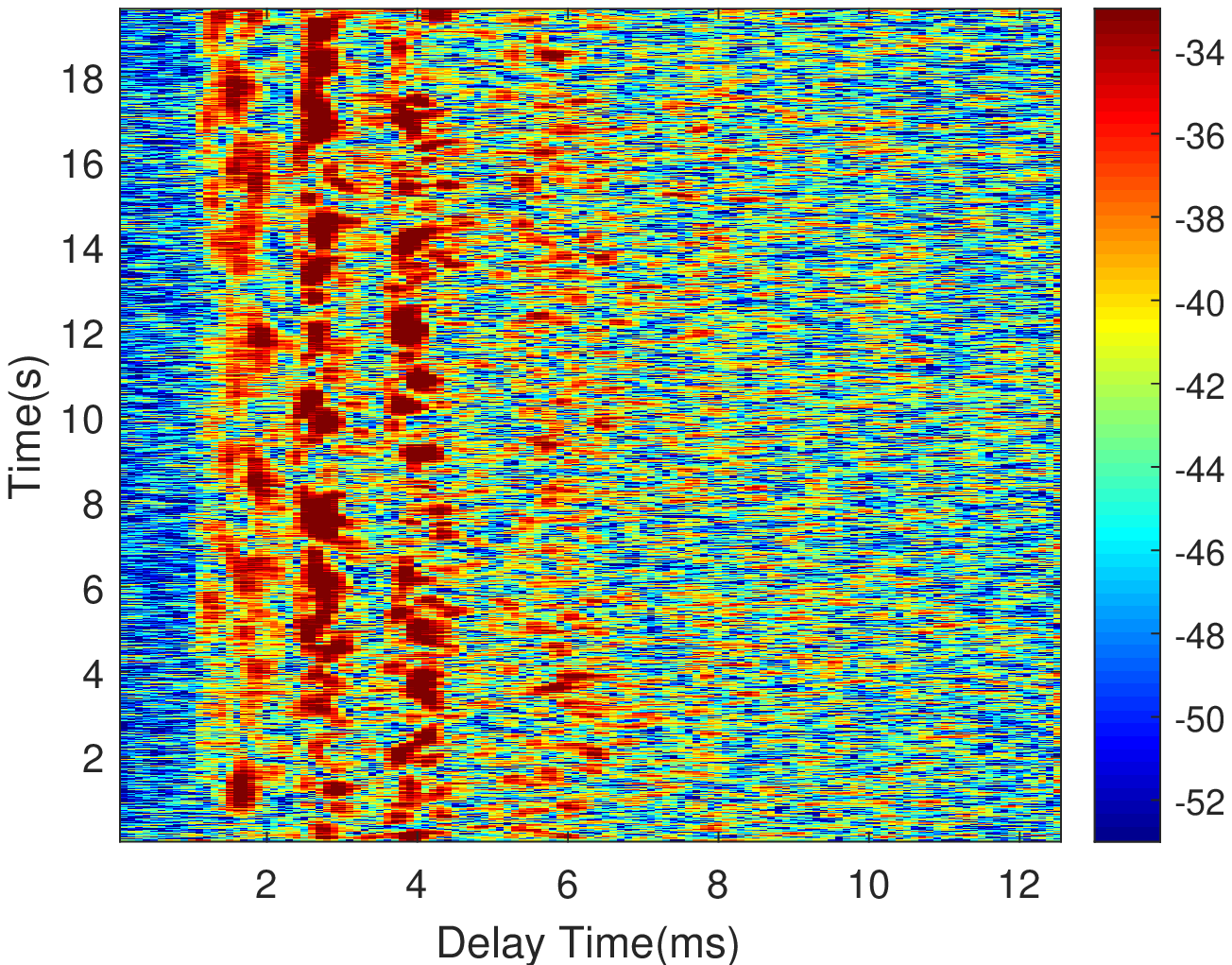}
	}
	\caption{Measured CIRs for the AUVFest07 data under the calm sea (a) and the rough sea (b)\cite{Huang2015Improving}.}
	\label{CIRs}
\end{figure*}

Therefore, the relationship between $\hat{\mathbf{Z}}_{\mathrm f}(n)$, $\hat{\mathbf{Z}}_{\mathrm b}(n)$ and $\tilde{\mathbf{Z}}(n)$  is as follows:
\begin{equation}
\tilde{\mathbf{M}}^{-1}(n) \tilde{\mathbf{Z}}(n) =\mathbf{M}_{\mathrm f}^{-1}(n)  \hat{\mathbf{Z}}_{\mathrm f}(n)+\mathbf{M}_{\mathrm b} ^{-1}(n) \hat{\mathbf{Z}}_{\mathrm b}(n),
\end{equation}
\begin{equation}
\tilde{\mathbf{M}}^{-1}(n) =\mathbf{M}_{\mathrm f}^{-1}(n) +\mathbf{M}_{\mathrm b}^{-1}(n)-\mathbf{R}_{\mathbf{z}}^{-1}(n),
\end{equation}
where $\mathbf{R}_{\mathbf{z}}$ can be omitted in order to simplify the problem.  $\hat{\mathbf{Z}}_{\mathrm f}(n)=\hat{\mathbf{Z}}(n,n)$ as in (\ref{zn}). For LMMSE, the error covariance $\tilde{\mathbf{M}}(n)$, $\mathbf{M}_{\mathrm f}(n)$ and $\mathbf{M}_{\mathrm b} (n)$ are also MSEs, and therefore, $\mathbf{M}_{\mathrm f} (n)=\mathbf{K}(n, n)$, and $\mathbf{M}_{\mathrm b} (n)=\mathbf{K}_{\mathrm b}(n,n)$.
Then,  for forward-backward Kalman filtering, a refined estimate $\tilde{\mathbf{Z}}(n)$ at time $n$ is as follows:
\begin{equation}\label{Mbf}
\tilde{\mathbf{M}}(n)=\left(\mathbf{K}^{-1}(n, n)+\mathbf{K}_{\mathrm b}^{-1}(n , n)\right)^{-1},
\end{equation}
\begin{equation}\label{bftra}
\tilde{\mathbf{Z}}(n)=\tilde{\mathbf{M}}(n)\left(\mathbf{K}^{-1}(n,n) \hat{\mathbf{Z}}(n,n)+\mathbf{K}_{\mathrm b}^{-1}(n , n) \hat{\mathbf{Z}}_{\mathrm b}(n,n)\right).
\end{equation}

Then, $\tilde{\mathbf{Z}}(n)$ in \eqref{bftra} after the forward-backward Kalman filter contains the final tracked channel components. 
 $\mathbf{\hat h}_{\mathrm{DFB-ASRMAE}}(n) = {\mathbf{Q}_r}(n)\tilde{\mathbf{Z}}(n)$, and is the tracked channel with the proposed DFB-ASRMAE algorithm. 
The detailed steps of DFB-ASRMAE algorithm are given in Table 1, and the algorithm has a complexity on the order of $O(\max(6Kr,10(rp)^3))$. Note that the ASRMAE algorithm has a computation complexity on the order of $O(\max(6Kr,3(rp)^3))$  \cite{Huang2015Improving}.  To obtain \eqref{bftra}, we need parameters from forward filtering and backward filtering. Hence, in the algorithm, we carry out forward filtering, backward filtering, and then the optimal joint estimation, as shown  in Table 1.

\begin{figure*}[h]
\xdef\xfigwd{\textwidth}
{	\centering
    \subfigure[]{
    	\centering
    	\includegraphics[scale=0.65]{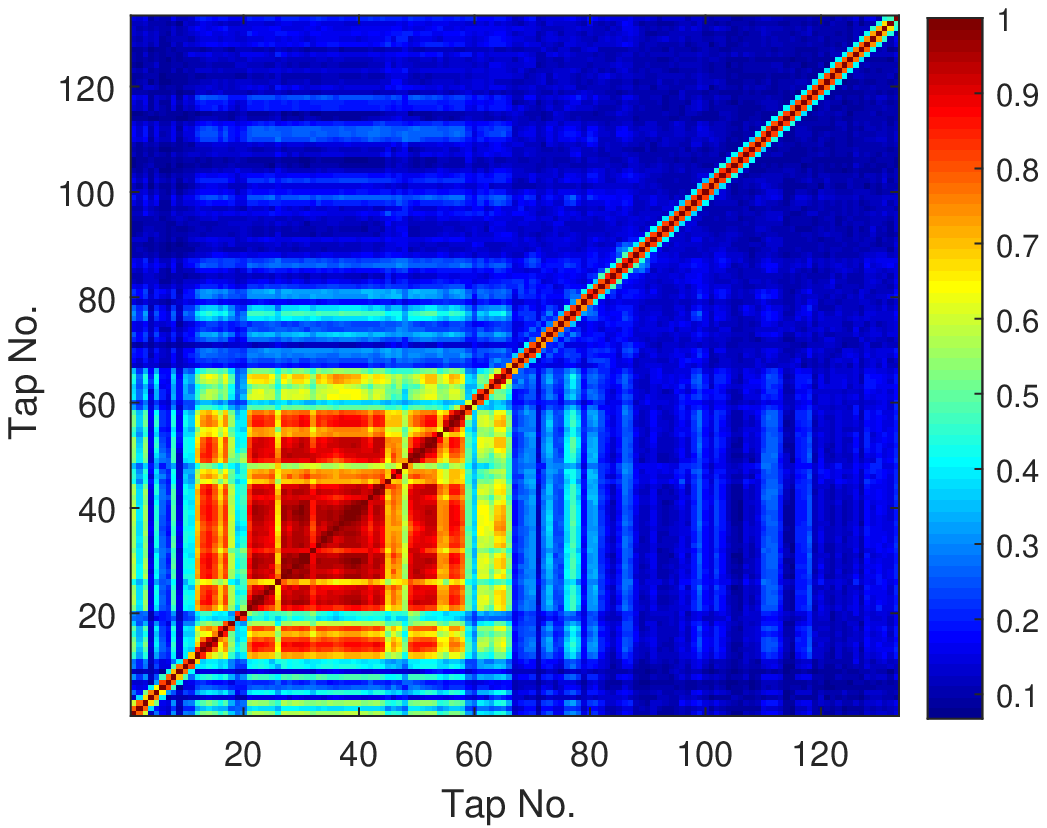}
    }
    \subfigure[]{
    	\centering
    	\includegraphics[scale=0.65]{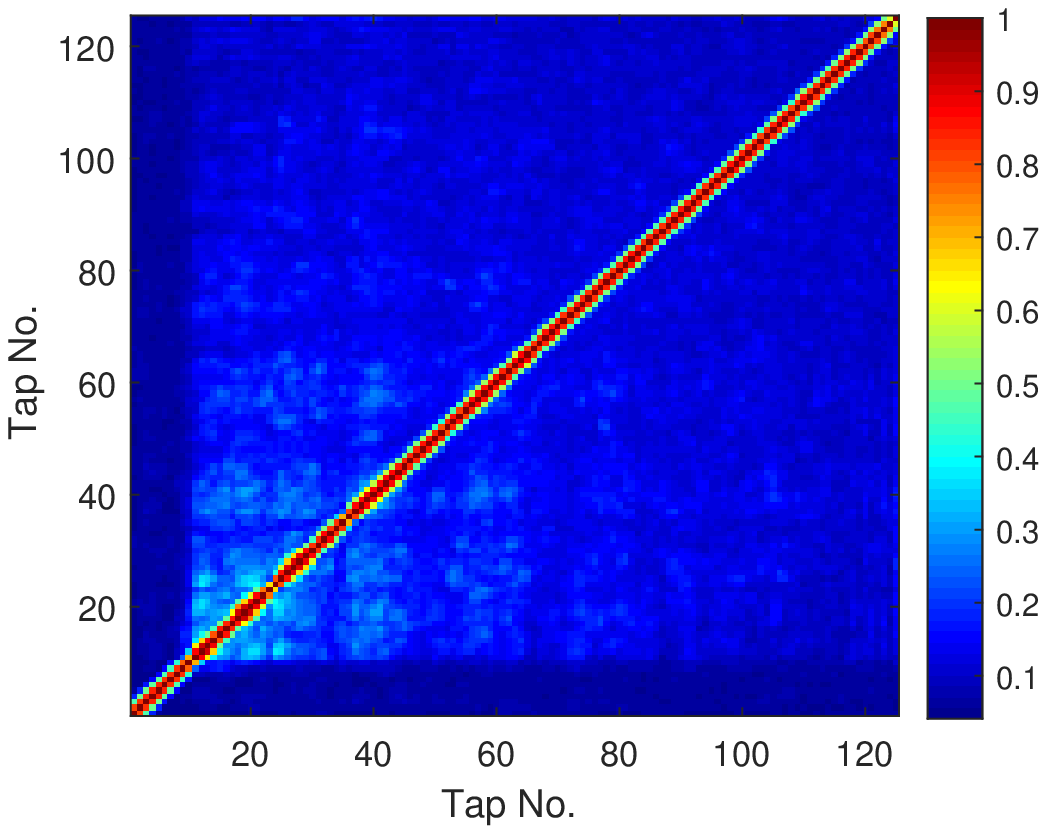}
    }
   \caption{The cross-path coherence estimated for AUVFest07 in the calm sea  (a) and the rough sea (b). The Tap No. refers to the $k$-th delay tap in $h_k(n)$ at time $n$.}
	\label{CIRtimes}}
\end{figure*}

\section{Experimental Results}
The 2007 Autonomous Underwater Vehicle Festival
(AUVFest07) experiment acoustic communication data previously studied in \cite{yang2012properties, Huang2014Model, Huang2015Improving} are used to verify our proposed channel tracking algorithm.
The experiment was conducted in 20 m of coastal water under relatively calm and rough sea conditions. 
Since the source and receiver are mounted on the bottom of a rigid body,  we do not need to consider the signal fluctuations due to the source or receiver's movement.  The measured CIRs are shown in Fig. \ref{CIRs}. The CIRs under both rough sea and calm sea are time variant due to internal waves.
Furthermore, the CIRs in the rough sea are more rapidly time-varying than those in the calm sea. For both sea conditions, three to four dominant paths are clearly shown in Fig. \ref{CIRs}, and the delay for each path does not belong to a single delay tap but rather spreads to multiple adjacent taps. This means that the channel taps are correlated with respect to delay for both calm and rough seas.

The cross-path coherence \cite{yang2012properties}  can be given by 
\begin{equation}
p_h[j,k]=\frac{E\left[h^{*}_j(n) h_k(n)\right]}{\sqrt{E\left[\left|h_j(n)\right|^{2}\right] E\left[\left|h_k(n)\right|^{2}\right]}},
\end{equation} 
 where  $E\left[\left|h_k(n)\right|^{2}\right]=\frac{1}{N_{p}} \sum_{n=1}^{N_{p}} \hat{h}_{k}^*(n) \hat{h}_{k}(n)$, and $\hat h_k(n)$ corresponds to the estimated $k$-th delay tap of channel at time $n$. 
Fig. \ref{CIRtimes} shows the cross-path coherence in different sea conditions. Even in the rough sea, some off-diagonal elements still have magnitudes comparable to those of the diagonal elements. This means that the channel taps are correlated in both calm and rough sea conditions.

Due to the above correlation characteristic of the channel, we can use EVD  to obtain eigenvalues $\mathbf{\Lambda}(n)$.  The normalized eigenvalue spectra for the two different sea conditions (in Fig. \ref{CIRs}) are shown in Fig. \ref{eigenvalues}.   The eigenvalue drops  rapidly from $k=1$ to 10 and becomes stable between  11 to 20. This means that the rank of the channel subspace can be chosen as $11\leq r \leq 20$. This is much smaller than the dimension of the CIR taps ($K \geq 100$).  Moreover, the curves showing the decrease in the eigenvalues in the calm seas and rough seas are similar. Thus, we can choose the same $r$ in different sea conditions.

\begin{figure*}[h]
 	\centering
 	\subfigure[]{
 		\centering
 		\includegraphics[scale=0.65]{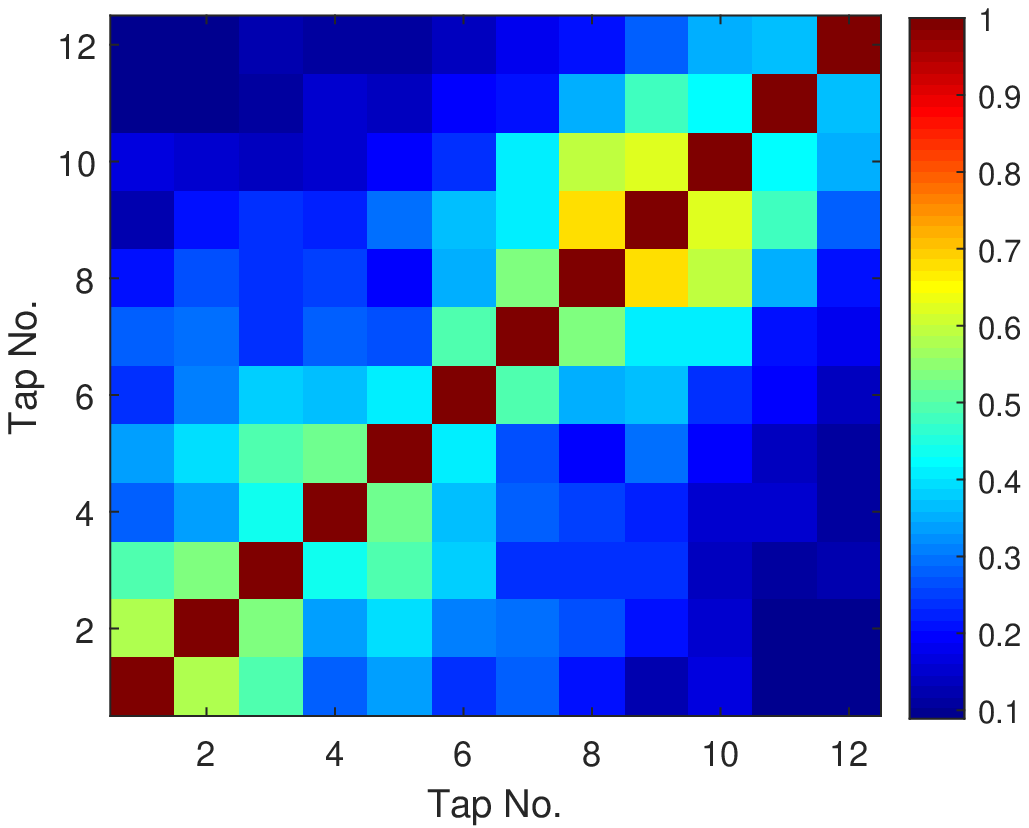}
 	}
 	\subfigure[]{
 		\centering
 		\includegraphics[scale=0.65]{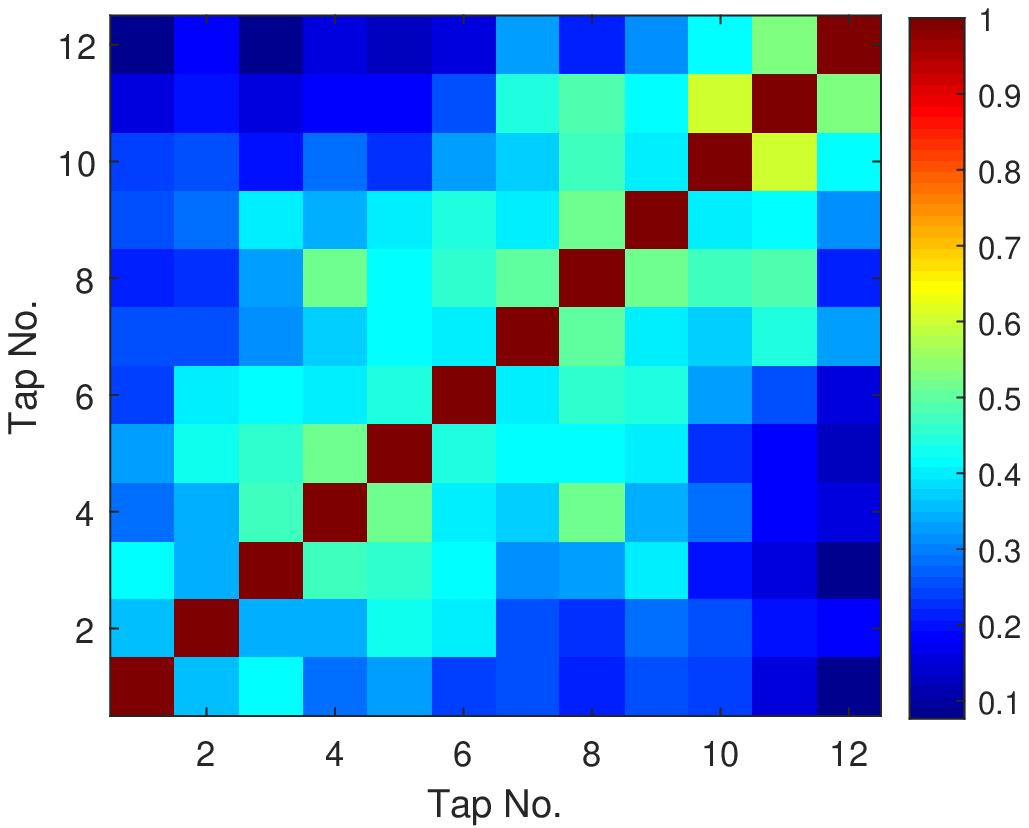}
 	}
 	\centering
 	\caption{Cross-path coherence of the channel components with $p=1$ and $r=12$ for the calm sea (a)  and the rough sea (b). Tap No. refers to $i$-th tap in $z_i(n)$ at time $n$.}
 	\label{CPCosigcom}
 \end{figure*}

We define a cross-path coherence for the channel principle components as follows, which  is similar to the cross-path coherence of a channel:
\begin{equation}
\rho_z[i,j]=\frac{E\left[z^{*}_i(n) z_j(n)\right]}{\sqrt{E\left[\left|z_i(n)\right|^{2}\right] E\left[\left|z_j(n)\right|^{2}\right]}},
\end{equation}
where  $E\left[\left|z_i(n)\right|^{2}\right]=\frac{1}{N_{p}} \sum_{n=1}^{N_{p}} \hat{z}_{i}^*(n) \hat{z}_{i}(n)$, and $\hat z_i(n)$ corresponds to the estimated channel  component of the $i$-th tap at time $n$.  The cross-path coherence of the channel components  is shown in Fig. \ref{CPCosigcom}. We observe that diagonal elements contain most of the power. It is reasonable that  conventional channel tracking assumes
 that the channel components are generally uncorrelated for different taps.
However, some of the off-diagonal elements have magnitudes comparable to those of the diagonal elements both in calm and rough sea conditions (the cross-path covariance matrix is not a strict diagonal matrix). Therefore, to compensate for  the conventional assumption, we assume that the process noise is correlated in the space-time model.

\Figure[t!][scale=0.5]{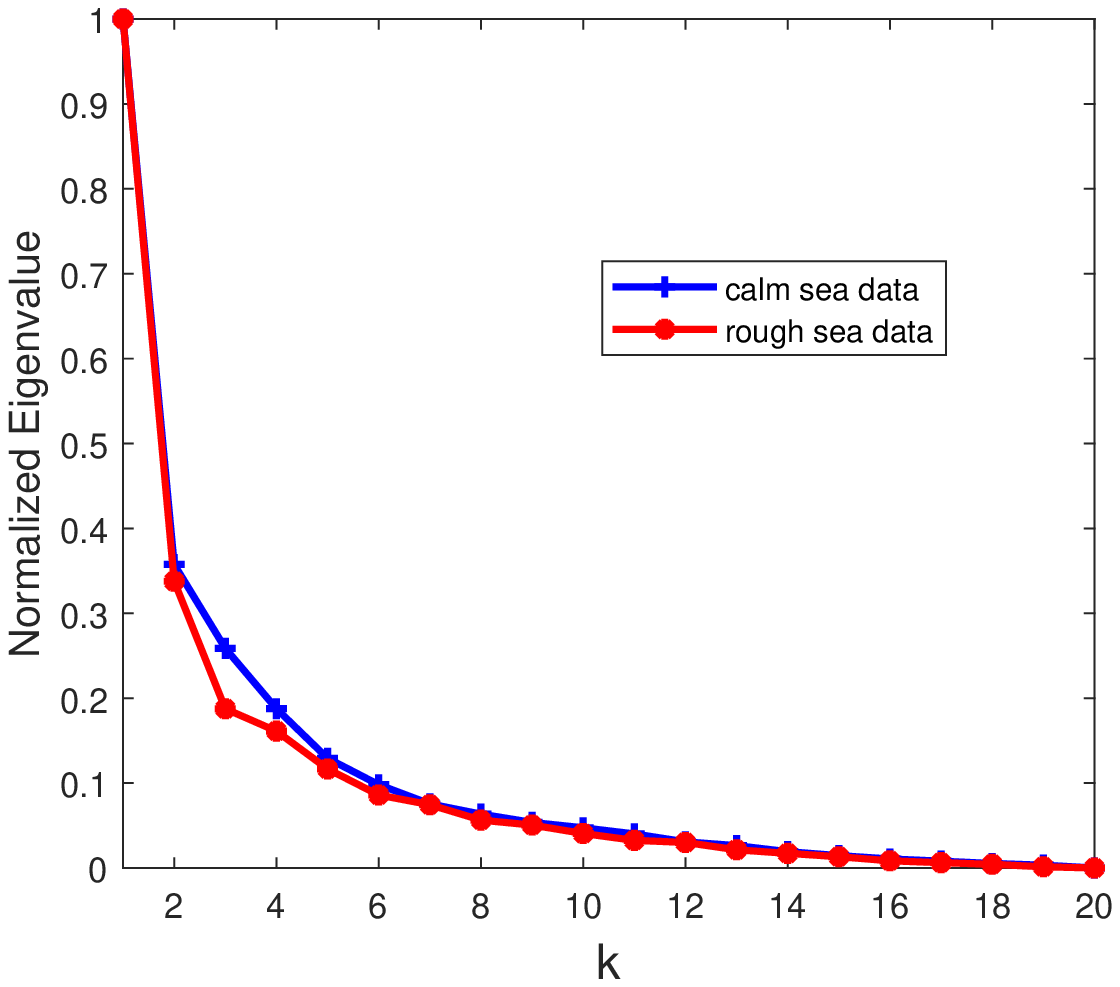}
{The normalized eigenvalues ($k$th diagonal element in $\mathbf{\Lambda}(n)$) of the channel for the calm sea and the rough sea.
	\label{eigenvalues}}

\begin{figure*}[h]
	\centering
	\subfigure[]{
		\centering
		\includegraphics[scale=0.45]{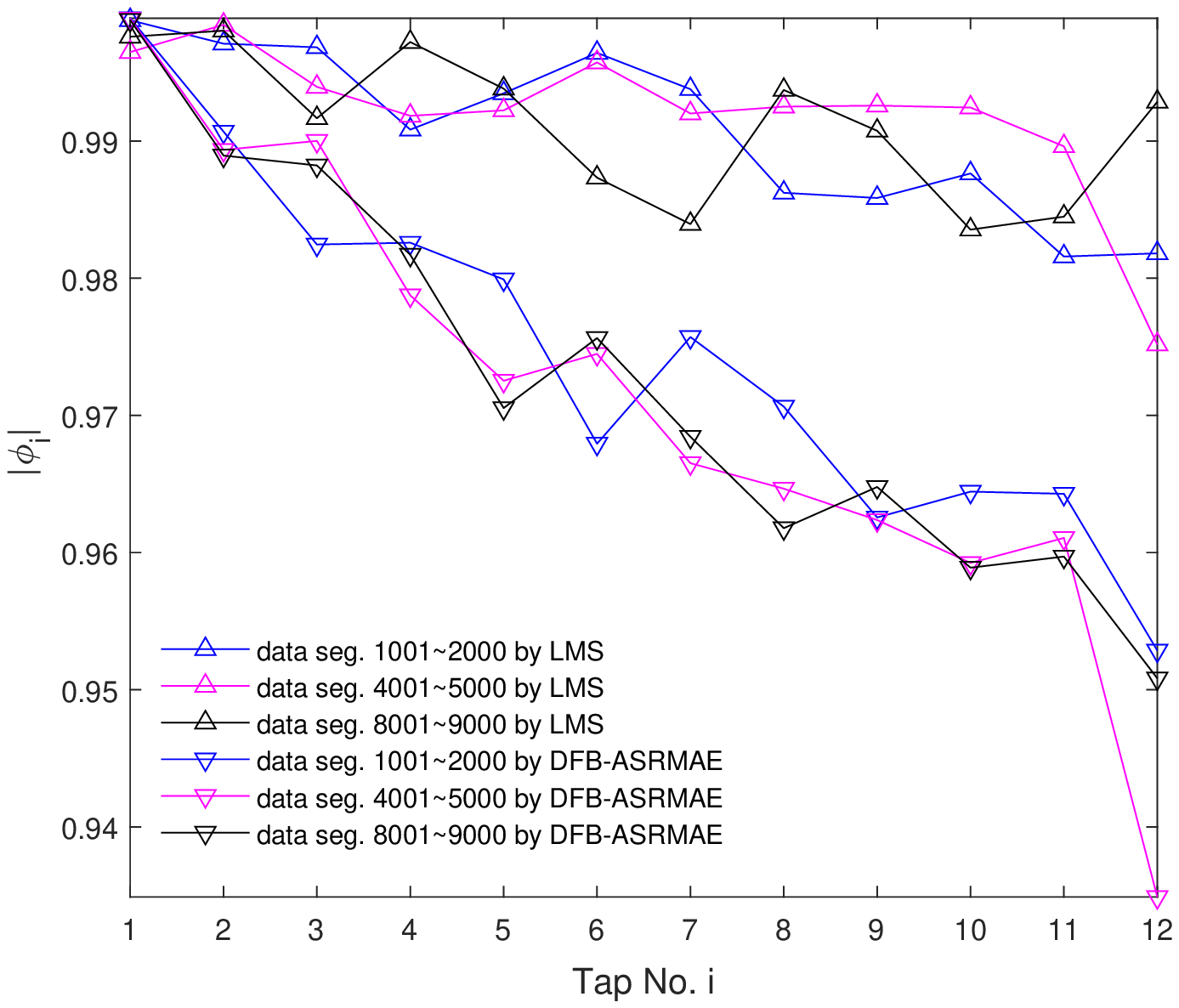}
	}
	\subfigure[]{
		\centering
		\includegraphics[scale=0.45]{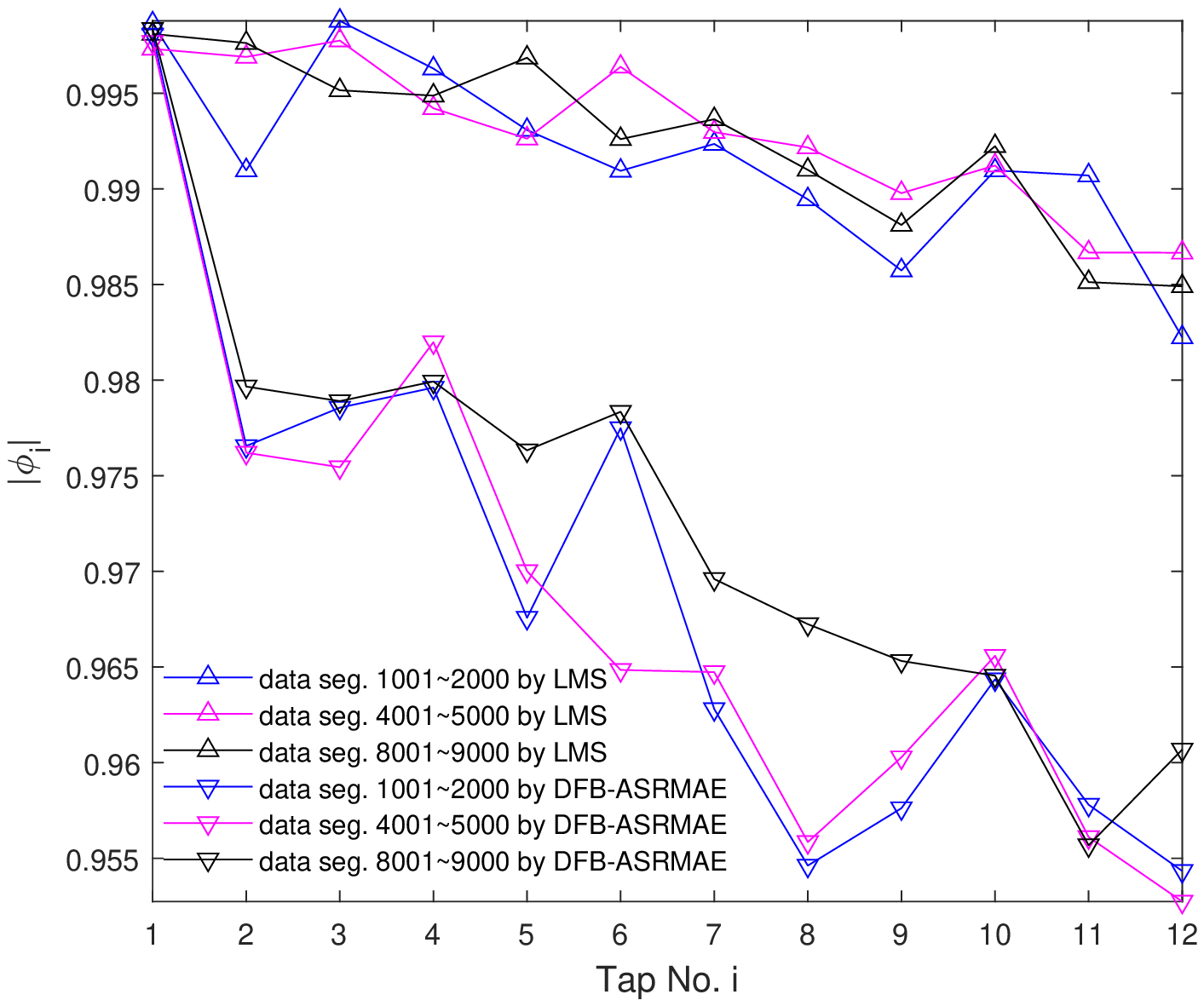}
	}
	\centering
		\caption{State transition coefficient $|\phi_i|$ for tap $i=1, \cdots, 12,$  with $p=1$ and $r=12$ where ${z}_i$ were estimated by LMS and DFB-ASRMAE algorithms with different data segments for the calm sea (a)  and the rough sea (b).}
	\label{statranscoefficients}
\end{figure*}

To analyze the state transition matrix,
here, we use a low-order state-space AR model with $p=1$, and then, the state transition matrix becomes a diagonal matrix. 
Figs. \ref{statranscoefficients}(a)-(b) show the state transition coefficients in the diagonal matrix obtained by the conventional method based on LMS as in Section III. A and the proposed DFB-ASRMAE under different sea conditions.
The dynamically updated state transition coefficients obtained by DFB-ASRMAE are different from the conventional state transition coefficients.
Moreover, the dynamically updated state transition coefficients in rough sea show more changes with different data time-segments. This observation suggests that  the state transition coefficients are time variant at a certain level and that
dynamically updating the state transition matrix can improve channel tracking in the rough sea more than in the calm sea.
\begin{figure*}[h]
	\centering
	\subfigure[]{
		\centering
		\includegraphics[scale=0.45]{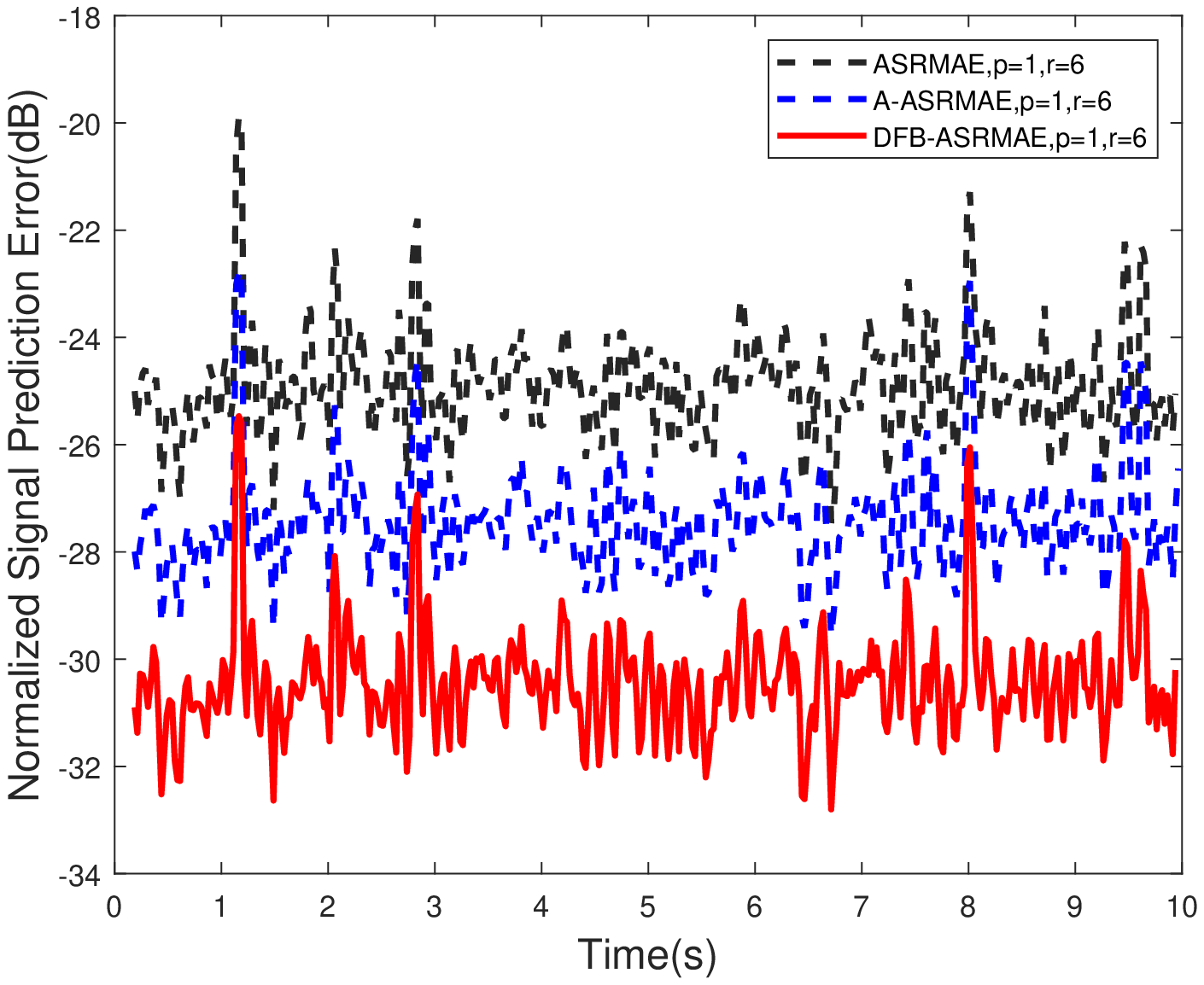}
	}
	\subfigure[]{
		\centering
		\includegraphics[scale=0.45]{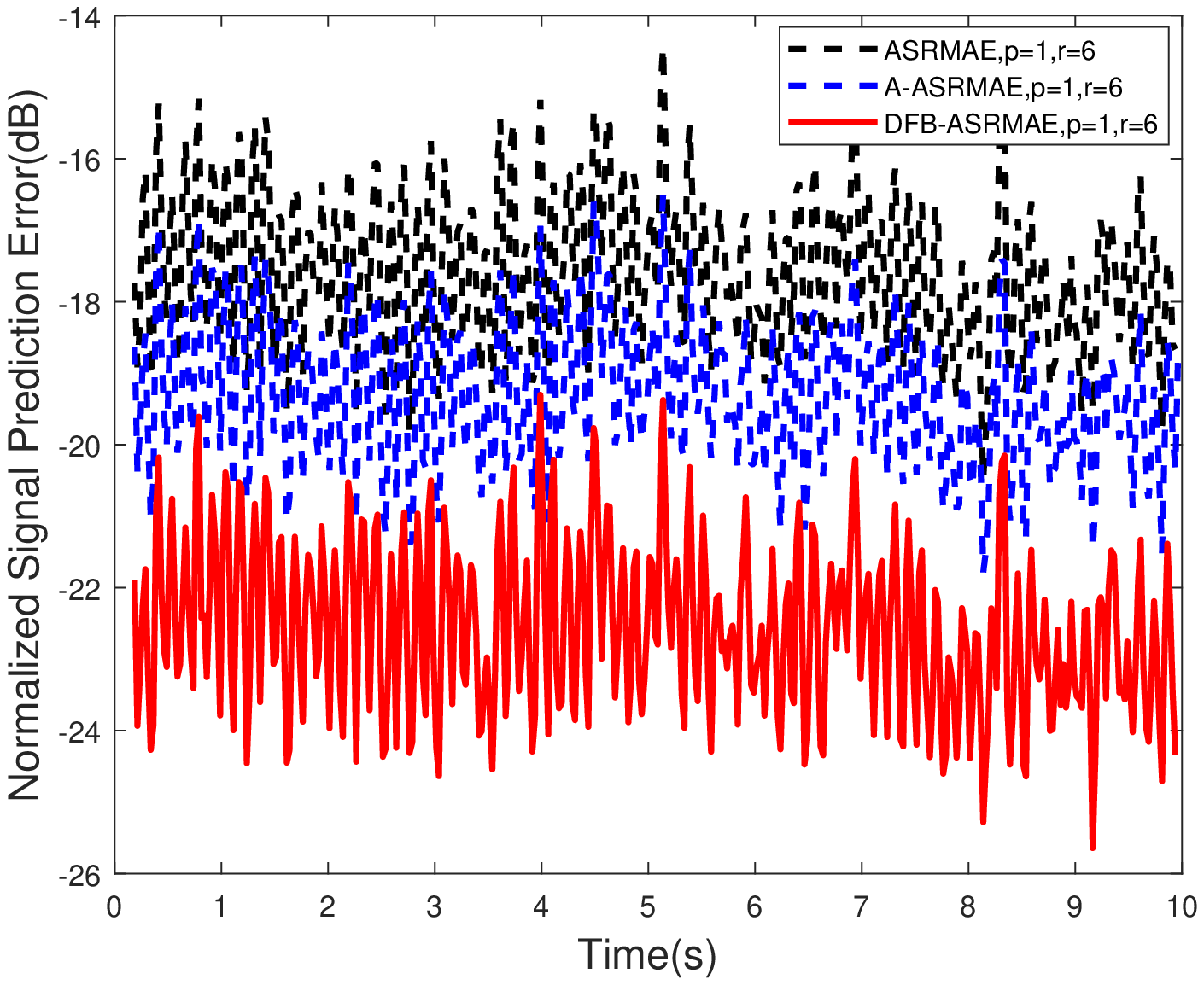}
	}
	\centering
	\caption{ Comparison of normalized signal prediction error as a function of time using various channel tracking algorithms for AUVFest07 in calm sea (a)  and  in rough sea (b). The mean normalized signal prediction errors in the calm sea are -25.1 dB for ASRMAE, -27.4 dB for A-ASRMAE, and -30.4 dB for DFB-ASRMAE, as marked on the right vertical axis in (a). In the rough sea (b), the mean normalized signal prediction errors are -17.6 dB for ASRMAE, -19.3 dB for A-ASRMAE, and -22.6 dB for DFB-ASRMAE.}
	\label{trackingalgorithms}
\end{figure*}
\Figure[t!][scale=0.6]{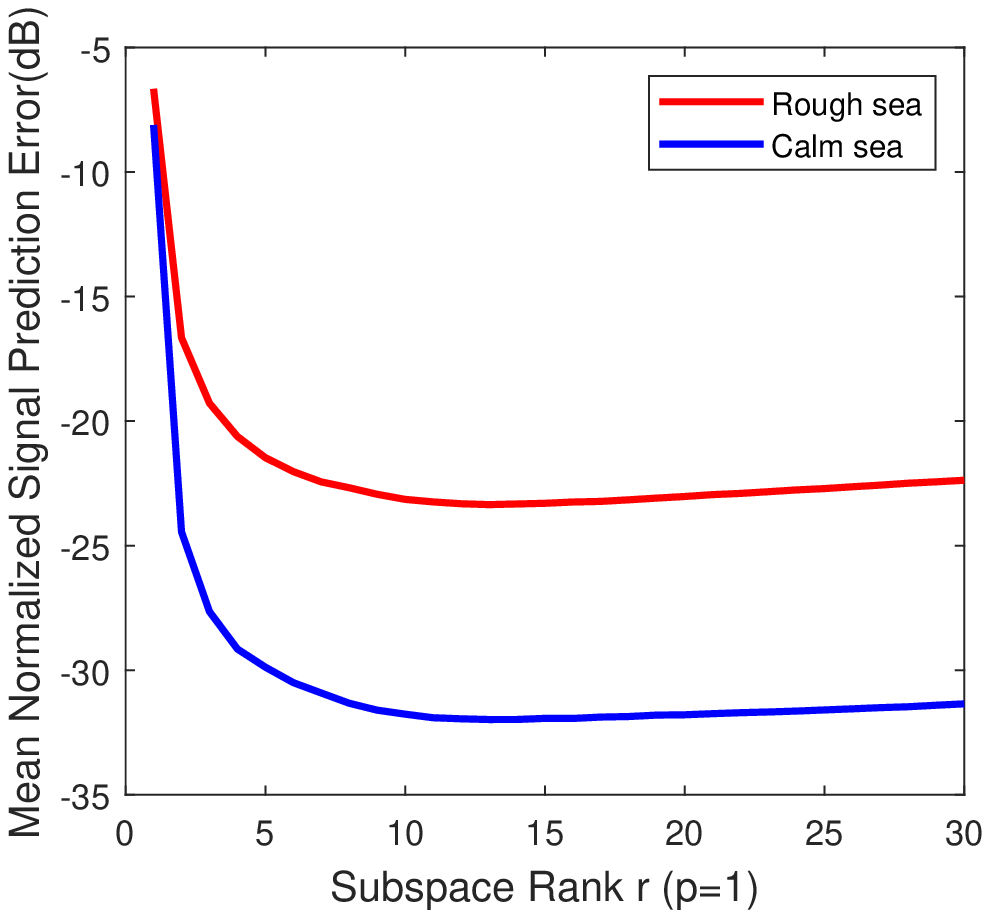}
{Mean normalized signal prediction error  by DFB-ASRMAE  as a function of $r$ for the AUVFest07 data.
	\label{Ranking}}
\begin{figure*}[h]
	\centering
	\subfigure[]{
		\centering
		\includegraphics[scale=0.45]{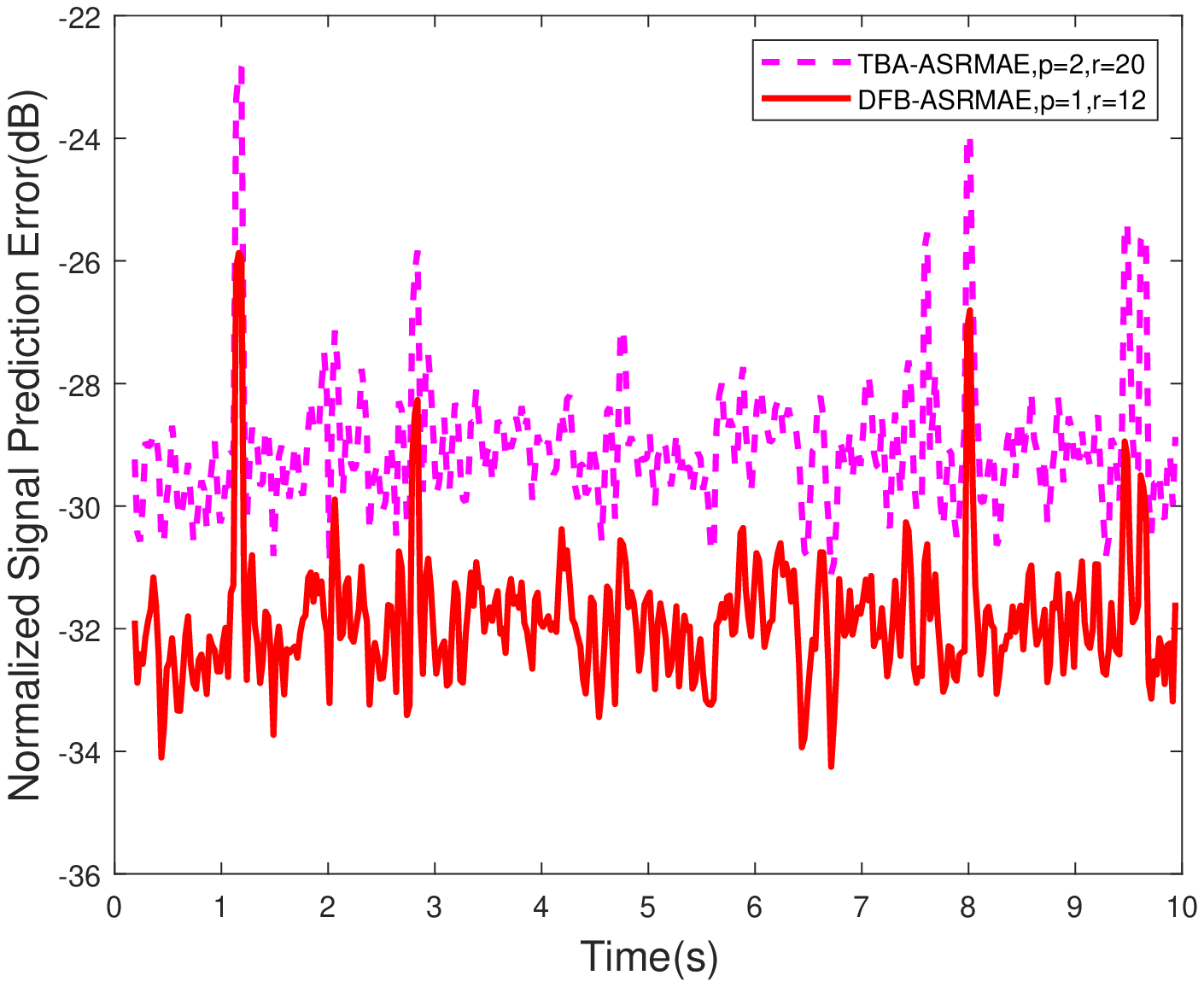}
	}
	\subfigure[]{
		\centering
		\includegraphics[scale=0.45]{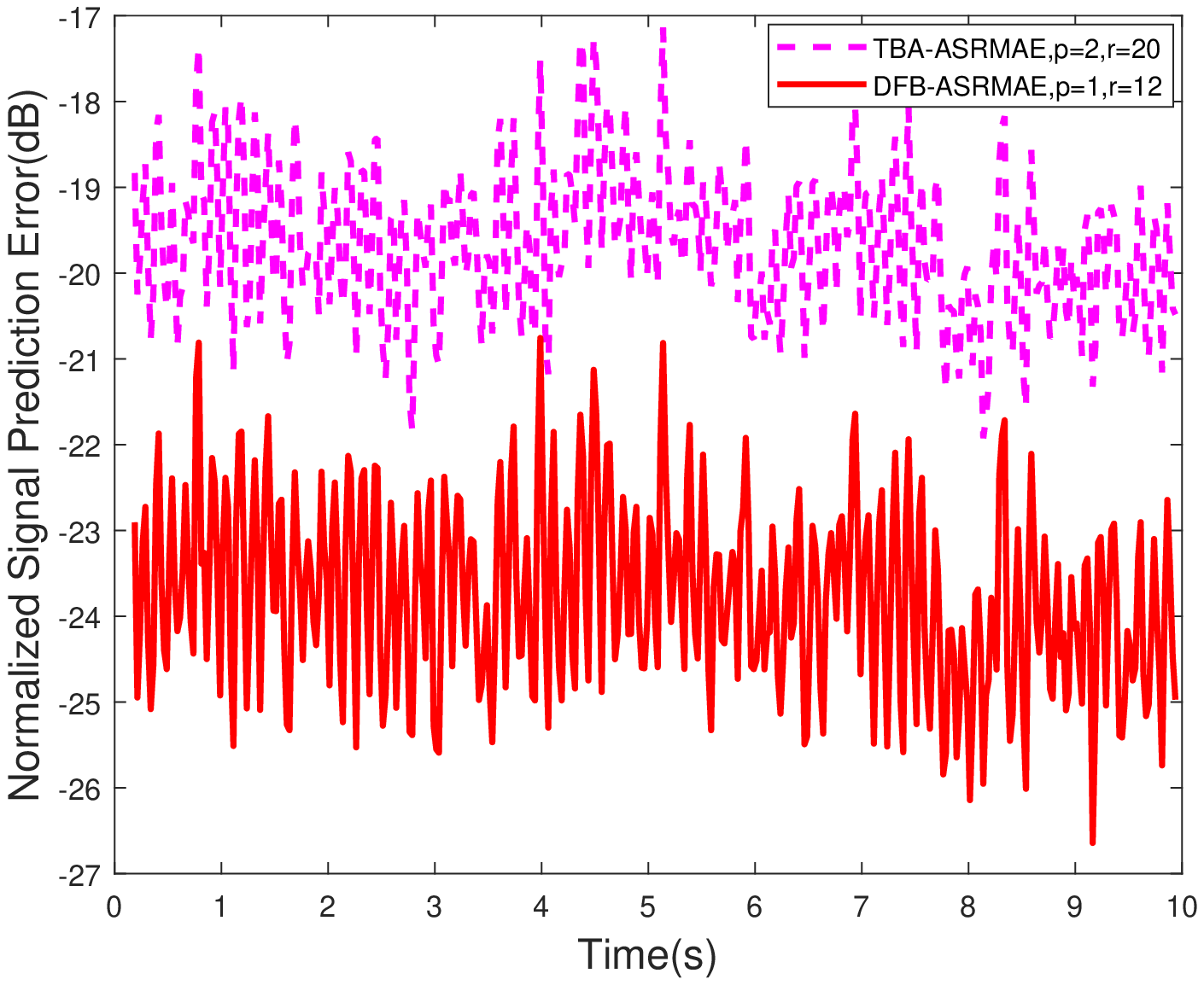}
	}
	\centering
	\caption{Comparison of the normalized signal prediction error as a function of time using various channel tracking algorithms for AUVFest07 in calm sea  (a) and  in rough sea (b). The mean normalized signal prediction errors for the calm sea data are -29.0 dB for TBA-ASRMAE ($p=2, r=20$) and -31.9 dB for DFB-ASRMAE ($p=1, r=12$). For the rough sea (b), the mean normalized signal prediction errors for the rough sea data are 19.7 dB for TBA-ASRMAE ($p=2, r=20$) and -23.7 dB for DFB-ASRMAE ($p=1, r=12$).}
	\label{TBA-DFB}
\end{figure*}

To compare the conventional methods in \cite{Huang2014Model, Huang2015Improving, wang2019training}, we continue to use the normalize  signal prediction error  as $|\xi(n)|^{2} / E|r(n)|^{2}$.

Figs. \ref{trackingalgorithms} (a) and (b) show the normalized signal prediction error tracked by ASRMAE \cite{Huang2015Improving}, A-ASRMAE \cite{wang2019training} and DFB-ASRMAE, separately, for the calm sea  and rough sea conditions with the AUVFest07 data. 
To compare the performance of different algorithms, we set the same  AR order $p=1$ and the same rank $r=6$ as in \cite{Huang2015Improving, wang2019training}. The mean normalized signal prediction errors for the calm sea data are -25.1 dB for ASRMAE, -27.4 dB for A-ASRMAE, and -30.4 dB for DFB-ASRMAE, as shown in Fig. \ref{trackingalgorithms}(a). For the rough sea data, the mean normalized signal prediction errors are -17.6 dB for ASRMAE, -19.3 dB for A-ASRMAE and -22.6 dB for DFB-ASRMAE, as shown in Fig. \ref{trackingalgorithms}(b).

The mean normalized signal prediction errors of A-ASRMAE and DFB-ASRMAE are lower than those of ASRMAE  for both the calm sea and rough sea conditions. A-ASRMAE yields a mean 2.3 dB improvement in the calm sea and 1.7 dB improvement in the rough sea in the signal prediction error compared with the ASRMAE algorithm since it introduces an adaptive procedure to compensate for the mismatch of the tracking model.
For the DFB-ASRMAE algorithm,  the mean signal prediction error is the lowest among the three algorithms: under the setting $p=1$ and $r=6$, it is 5.3 dB and 5.0 dB lower than those obtained by ASRMAE for the calm sea conditions and the rough sea conditions, respectively.  DFB-ASRMAE adopts a dynamic time-variant space-time model with correlated tolerance. Therefore, it can improve the channel tracking performance in principle. Moreover, the forward-backward filtering can further help to track the channel with this time-variant model.

Furthermore, we discuss the parameter setting.
Since higher-order $p$ can gives rise to a high complexity of DFB-ASRMAE, we set $p=1$ in this work.
We provide the tracking results of the DFB-ASRMAE algorithm with changing  $r$ under $p=1$   in Fig. \ref{Ranking}.
The results show that the mean normalized prediction error reaches the lowest value at approximately $r=12$ with the proposed DFB-ASRMAE for both the calm sea data and rough sea data. Based on the examination of  Figs. \ref{Ranking} and \ref{eigenvalues}, we can reach the same conclusion that it is reasonable to set the same rank $r$ for different sea conditions.

Therefore, we set $p=1$ and $r=12$ for DFB-ASRMAE and compare it with TBA-ASRMAE \cite{wang2019training} with the same AUVFest07 data.
In TBA-ASRMAE, all of the parameters are set according to its training procedure, and the AR order and subspace rank are  $p=2$ and $r=20$, respectively, and are  the optimal parameters for TBA-ASRMAE in both calm seas and rough seas.
Both show performance superior to those of ASRMAE and  A-ASRMAE (see Fig. \ref{trackingalgorithms}) with $p=1$ and $r=6$.
DFB-ASRMAE  yields a 2.9 dB improvement in the signal error prediction error in the calm sea and a 4.0 dB improvement in the rough sea compared with TBA-ASRMAE, which has a higher AR order and more subspaces.
Furthermore, the DFB-ASRMAE  shows more improvement in the rough sea than in the calm sea.
Even though the rough channel changes strongly over time,
the dynamical time-variant space-time model with correlated tolerance and a forward-backward Kalman filter in DFB-ASRMAE can work well together to track the channel's rapid changes.

\section{CONCLUSION}

In this work,  we consider the model mismatch problem for model-based channel tracking. The model is based on channel physics  in that the underwater acoustic channel is correlated. In the model, the channel components are assumed to be uncorrelated with a time-invariant transaction after decorrelation. However, this assumption does not always hold for the real underwater acoustic channel.  Therefore, we propose a dynamic state-space model that is more tolerant to the rapid time-varying channel for channel tracking. Furthermore, a forward-backward Kalman filter is combined with the dynamic state-space model,  further improving the tracking performance. The proposed DFB-ASRMAE with only a 1-order AR model can decrease the normalized signal prediction error significantly with experimental data only.
\appendices
\section*{Acknowledgment}
The authors would like to thank  Prof T. C. Yang and Dr. X Han for providing the AUVFest07 data.

\ifCLASSOPTIONcaptionsoff
  \newpage
\fi

\begin{IEEEbiography}[{\includegraphics[width=1in,height=1.25in,clip,keepaspectratio]{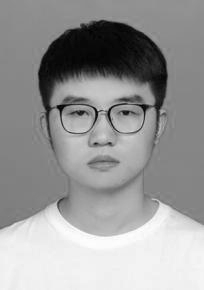}}]{Qihang Huang}  received his B.S. degree in electronic and information engineering from Harbin Institute of Technology, Harbin, China, in 2019. He is currently pursuing his M.S. degree in electronics and communication engineering at Harbin Institute of Technology, Shenzhen, China. His research interests include underwater acoustic communication, algorithm design and channel tracking.
\end{IEEEbiography}

\begin{IEEEbiography}[{\includegraphics[width=1in,height=1.25in,clip,keepaspectratio]{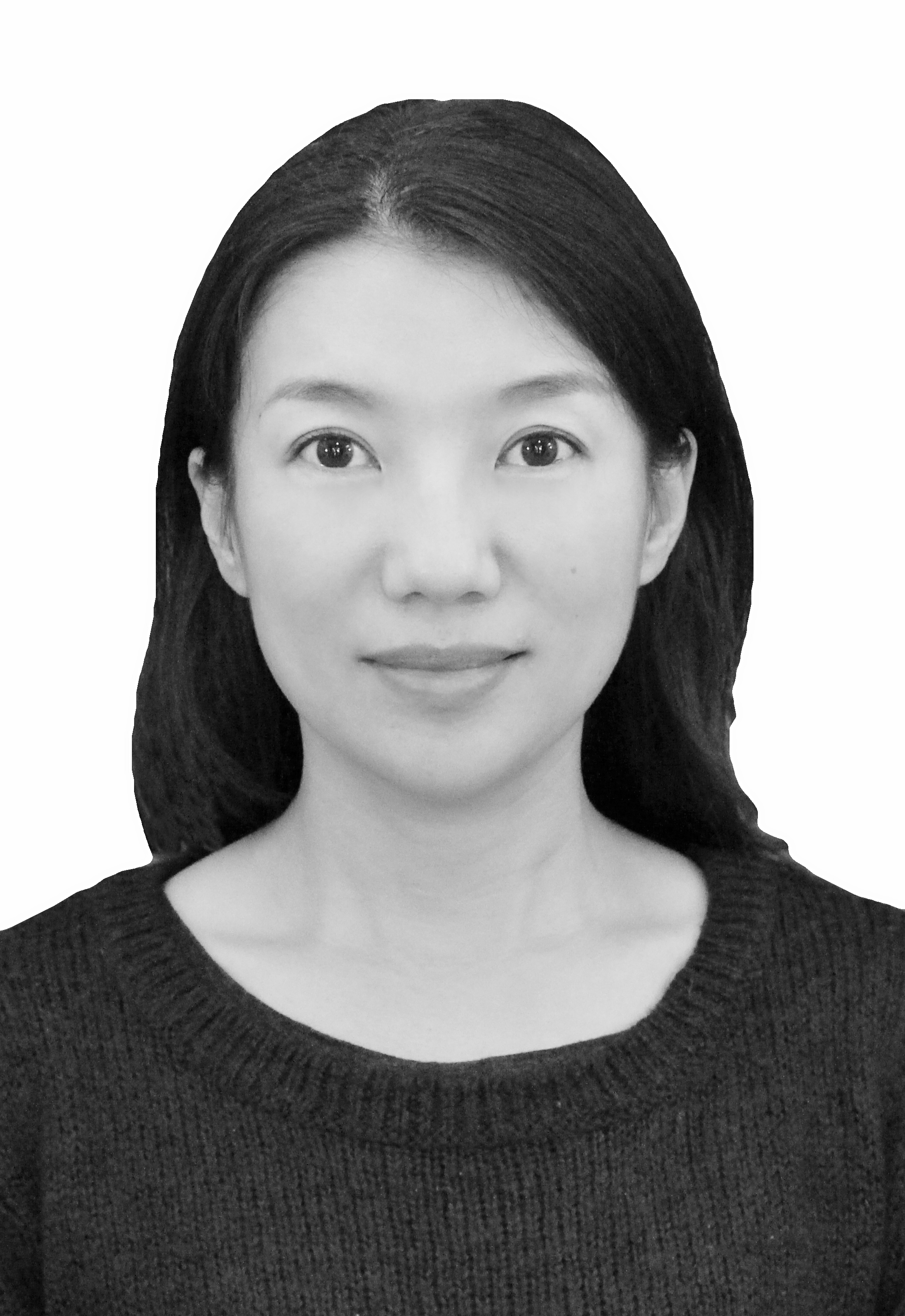}}]{Wei Li}  
 (M'11) received her B.S. degree in electric and information engineering from Dalian University of Technology, Dalian, China, in 2004 and Ph.D. degrees in signal and information processing from Institute Of Acoustics, Chinese Academy Of Sciences, Beijing, China, in 2009. She visited the University of Connecticut, Storrs, CT, USA, from September 2015 to October 2016. She is an Associate Professor with the school of electric and information engineering at Harbin institute of Technology (Shenzhen). Her research interests lie in the areas of underwater acoustic communication, underwater acoustic detection and estimation, statistical signal processing, and wireless communications.
\end{IEEEbiography}

\begin{IEEEbiography}[{\includegraphics[width=1in,height=1.25in,clip,keepaspectratio]{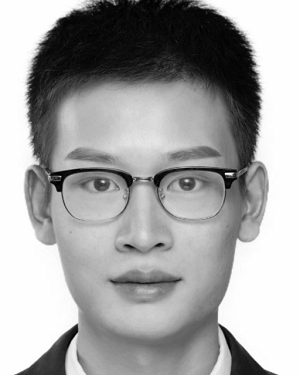}}]{Weicheng Zhan} received his B.S. degree in communication engineering from the Harbin Institute of Technology, Weihai, China, in 2019, He is currently pursuing his M.S. degree in electronics and communication engineering at Harbin Institute of Technology, Shenzhen, China.
His research interests include hydroacoustic channel estimation and tracking, signal detection and estimation, and compression sensing.

\end{IEEEbiography}
\begin{IEEEbiography}[{\includegraphics[width=1in,height=1.25in,clip,keepaspectratio]{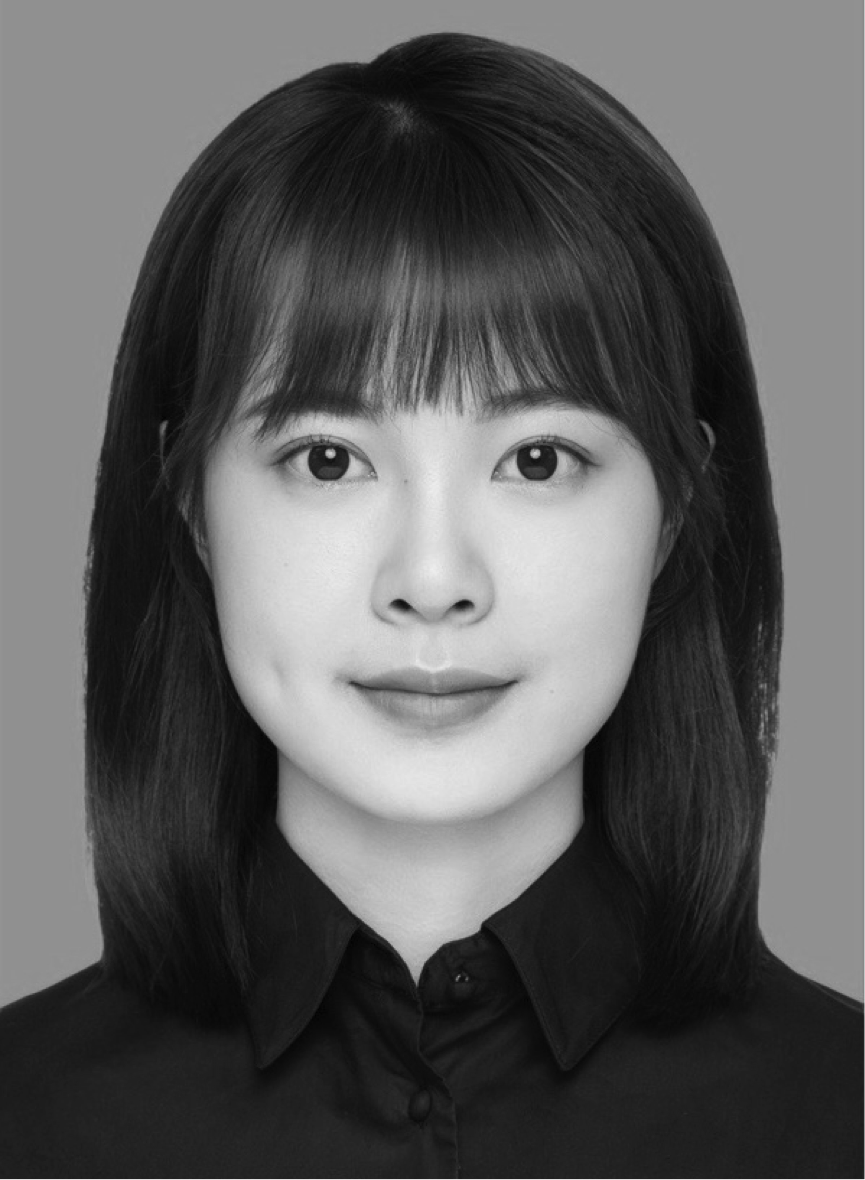}}]{Yuhang Wang} received her B.S. degree in communication engineering from Harbin Institute of Technology, Shenzhen, in 2020. She is currently pursuing her M.S. degree at Harbin Institute of Technology.
Her current research interests include underwater acoustic communication, underwater acoustic channel tracking.

\end{IEEEbiography}
\begin{IEEEbiography}[{\includegraphics[width=1in,height=1.25in,clip,keepaspectratio]{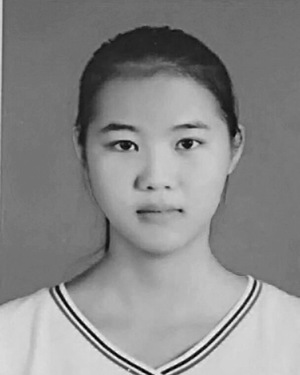}}]{Rongrong Guo} 
 received her B.S. degree in electronic and information engineering from Jiangxi Science and Technology Normal University, Nanchang, China, in 2020. She is currently pursuing her M.S. degree in electronic and information engineering at Harbin Institute of Technology, Shenzhen, China. Her current research interest is underwater acoustic communication.

\end{IEEEbiography}

\EOD

\end{document}